\documentclass{mn2e}
\usepackage{graphics}
\usepackage{epsfig}
\usepackage{color}
\usepackage{xspace}

\def\ni{\noindent}                                       
\def\etal{et\thinspace al.\ }                               

\newcommand{\Ha}{\ifmmode {\rm H}\alpha \else H$\alpha$\fi\xspace}
\newcommand{\Hb}{\ifmmode {\rm H}\beta \else H$\beta$\fi\xspace}
\newcommand{\Hg}{\ifmmode {\rm H}\gamma \else H$\gamma$\fi\xspace}
\newcommand{\Hd}{\ifmmode {\rm H}\delta \else H$\delta$\fi\xspace}

\newcommand{\Hii}{\ifmmode \rm{H}\,\textsc{ii} \else H~{\sc ii}\fi}

\newcommand{\nii}{\ifmmode [\rm{N}\,\textsc{ii}] \else [N~{\sc ii}]\fi}

\newcommand{\oi}{\ifmmode [\rm{O}\,\textsc{i}] \else [O~{\sc i}]\fi}

\newcommand{\neiii}{\ifmmode [\rm{Ne}\,\textsc{iii}] \else [Ne~{\sc iii}]\fi}
\newcommand{\hei}{\ifmmode [\rm{He}\,\textsc{i}] \else [He~{\sc i}]\fi}
\newcommand{\oii}{\ifmmode [\rm{O}\,\textsc{ii}] \else [O~{\sc ii}]\fi}

\newcommand{\oiii}{\ifmmode [\rm{O}\,\textsc{iii}] \else [O~{\sc iii}]\fi}

\newcommand{\sii}{\ifmmode [\rm{S}\,\textsc{ii}] \else [S~{\sc ii}]\fi}
\newcommand{\siii}{\ifmmode [\rm{S}\,\textsc{iii}] \else [S~{\sc iii}]\fi}

\title[Semi-empirical analysis of SDSS galaxies]
          {Semi-empirical analysis of SDSS galaxies: I.
Spectral synthesis method}

\author[Cid Fernandes et al]
         {Roberto Cid Fernandes$^{1}$\thanks{E-mail: cid@astro.ufsc.br},
	 Ab\'{\i}lio Mateus$^{2}$\thanks{E-mail: abilio@astro.iag.usp.br},
	 Laerte Sodr\'e Jr.$^{2}$\thanks{E-mail: laerte@astro.iag.usp.br},
	 Gra\.zyna Stasi\'nska $^{3}$\thanks{E-mail:
grazyna.stasinska@obspm.fr}
	 \newauthor
	 Jean M. Gomes$^{1}$\thanks{E-mail: jean@astro.ufsc.br}\\
	 $^{1}$Depto.\ de F\'{\i}sica - CFM - Universidade Federal de
Santa Catarina,
	 Florian\'opolis, SC, Brazil\\
	 $^{2}$Instituto de Astronomia, Geof\'{\i}sica e Ci\^encias
         Atmosf\'ericas, Universidade de S\~ao Paulo, S\~ao Paulo, SP,
         Brazil\\
	 $^{3}$LUTH, Observatoire de Meudon, 92195 Meudon Cedex, France}

\begin{document}

\maketitle

\begin{abstract}
The study of stellar populations in galaxies is entering a new era
with the availability of large and high quality databases of both
observed galactic spectra and state-of-the-art evolutionary synthesis
models. In this paper we investigate the power of spectral synthesis
as a mean to estimate physical properties of galaxies. Spectral
synthesis is nothing more than the decomposition of an observed
spectrum in terms of a superposition of a base of simple stellar
populations of various ages and metallicities, producing as output the
star-formation and chemical histories of a galaxy, its extinction and
velocity dispersion. Our implementation of this method uses the
Bruzual \& Charlot (2003) models and observed spectra in the
3650--8000 \AA\ range. The reliability of this approach is studied by
three different means: (1) simulations, (2) comparison with previous
work based on a different technique, and (3) analysis of the
consistency of results obtained for a sample of galaxies from the
Sloan Digital Sky Survey (SDSS).

We find that spectral synthesis provides reliable physical parameters
as long as one does not attempt a very detailed description of the
star-formation and chemical histories. Robust and physically
interesting parameters are obtained by combining the (individually
uncertain) strengths of each simple stellar population in the base. In
particular, we show that besides providing excellent fits to observed
galaxy spectra, this method is able to recover useful information on
the distributions of stellar ages and, more importantly, stellar
metallicities. Stellar masses, velocity dispersion and extinction are
also found to be accurately retrieved for realistic signal-to-noise
ratios.

We apply this synthesis method to a volume limited sample of 50362
galaxies from the SDSS Data Release 2, producing a catalog of stellar
population properties. Emission lines are also studied, their
measurement being performed after subtracting the computed starlight
spectrum from the observed one. A comparison with recent estimates of
both observed and physical properties of these galaxies obtained by
other groups shows good qualitative and quantitative agreement,
despite substantial differences in the method of analysis. The
confidence in the method is further strengthened by several empirical
and astrophysically reasonable correlations between synthesis results
and independent quantities.  For instance, we report the existence of
strong correlations between stellar and nebular metallicites, stellar
and nebular extinctions, mean stellar age and equivalent width of
H$\alpha$ and 4000 \AA\ break, and between stellar mass and velocity
dispersion.

\end{abstract}


\section{Introduction}

\label{sec:Introduction}

Galaxy spectra encode information on the age and metallicity
distributions of the constituent stars, which in turn reflect the
star-formation and chemical histories of the galaxies. Retrieving this
information from observational data in a reliable way is crucial for a
deeper understanding of galaxy formation and evolution.

The mapping of observed onto physical properties of galaxies has been
a major topic of research for over a generation of astronomers since
the pioneering works of Morgan (1956), Wood (1966) and Faber (1972) on
the one hand, and Tinsley (1968) and Spinrad \& Taylor (1972) on the
other. The first group of authors introduced the so-called empirical
population synthesis methods, which aim at reproducing a set of
observations of a given galaxy by means of a linear combination of
simpler systems of known characteristics, like individual stars or
chemically homogeneous and coeval groups of stars.  Bica (1988), Pelat
(1997), Cid Fernandes \etal (2001), Moultaka \etal (2004) are examples
of modern studies following this approach. The second group pioneered
the so-called evolutionary population synthesis methods, which compare
galaxy data with models that follow the time evolution of an entire
stellar system by combining libraries of evolutionary tracks and
stellar spectra with prescriptions for the initial mass function
(IMF), star formation and chemical histories. This approach has
enjoyed more widespread use in recent years, e.g., Arimoto \& Yoshii
(1987), Guiderdoni \& Rocca-Volmerange (1987), Bressan \etal (1994),
Fioc \& Rocca-Volmerange (1997), Vazdekis (1999), Bruzual \& Charlot
(2003, hereafter BC03), Le Borgne \etal (2004), among many others (see
Cardiel \etal 2003 for a large set of references).  In short,
empirical synthesis relies on nature for its basic ingredients,
whereas evolutionary synthesis relies mostly on models. However, since
models are made and calibrated to mimic nature, this difference is
gradually vanishing as models improve.

Besides different methodologies, there are also differences in what
type of data is actually modeled. Colours (e.g., Wood 1966),
absorption line equivalent widths or spectral indices (e.g., Worthey
1994; Kauffmann \etal 2003, hereafter K03) and emission features, both
stellar (Leitherer, Robert \& Heckman 1995; Schaerer \& Vacca 1998)
and nebular (Mas-Hesse \& Kunth 1991; Kewley \etal 2001), have all
been used in stellar population synthesis.  More recently, the full
spectral information has been incorporated in the modeling process,
both including (Charlot \& Longhetti 2001), and excluding emission
lines (Vazdekis \& Arimoto 1999; Reichardt, Jimenez \& Heavens 2001;
Cid Fernandes \etal 2004a, hereafter CF04).

Recovering the stellar content of a galaxy from its observed
integrated spectrum is not an easy task, as can be deduced from the
amount of work devoted to this topic over the past $\sim$ half
century.  The situation is however much more favorable nowadays. Huge
observational and theoretical efforts in the past few years have
produced large sets of high quality spectra of stars (e.g.,
Prugniel \& Soubiran 2001; Le Borgne \etal 2003; Bertone \etal 2004;
Gonz\'alez Delgado \etal 2004). These libraries are being implemented
in a new generation of evolutionary synthesis models, allowing the
prediction of galaxy spectra with an unprecedented level of detail
(Vazdekis 1999; BC03, Le Borgne \etal 2004).  At the same time, galaxy
spectra are now more abundant than ever (Loveday \etal 1996; York
\etal 2000). The Sloan Digital Sky Survey (SDSS), in particular, is
providing a homogeneous data base of hundreds of thousands of
galaxy spectra in the 3800--9200 \AA\ range, with a resolution of
$\lambda / \Delta \lambda \sim 1800$ (York \etal 2000; Stoughton \etal
2002; Abazajian \etal 2003, 2004).  This enormous amount of high
quality data will undoubtedly be at the heart of tremendous progress
in our understanding of galaxy constitution, formation and
evolution. Indeed, significant steps in this direction have recently
been made (K03; Brinchmann \etal 2004, Tremonti \etal 2004; Heavens
\etal 2004; Panter, Heavens \& Jimenez 2004).

In order to take advantage of the recent progress in evolutionary
synthesis to analyze data sets such as the SDSS, a methodology must be
set up to go from the observed spectra to physical properties of
galaxies. In this paper we discuss one possible method to achieve this
goal. The method is based on fitting an observed spectrum with a
linear combination of simple theoretical stellar populations (coeval
and chemically homogeneous) computed with evolutionary synthesis
models at the same spectral resolution as that of the SDSS.

Our goal here is to demonstrate that, besides providing excellent
starlight templates to aid emission line studies, spectral synthesis
recovers reliable stellar population properties out of galaxy spectra
of realistic quality. We show that this simple method provides robust
information on the stellar age ($t_\star$) and stellar metallicities
($Z_\star$) distributions, as well as on the extinction, velocity
dispersion and stellar mass. The ability to recover information on
$Z_\star$ is particularly welcome, given that stellar metallicities
are notoriously more difficult to assess than other properties.  In
order to reach this goal we follow: (1) {\it a priori} arguments,
based on simulations; (2) comparisons with independent work based on a
different method, and (3) an {\it a posteriori} empirical analysis of
the consistency of results obtained for a large sample of SDSS
galaxies. Other papers in this series on the Semi-Empirical Analysis
of Galaxies (SEAGal) will explore various astrophysical implications
of the results obtained with this method.

This paper is organized as follows. Section \ref{sec:Synthesis}
presents an overview of our synthesis method and simulations designed
to test it and evaluate the uncertainties involved. The discussion is
focused on how to use the synthesis to derive robust estimators of
physically interesting stellar population properties. Section
\ref{sec:Syntesis_SDSS} defines a volume limited sample of SDSS
galaxies and presents the results of the synthesis of their spectra,
along with measurements of emission lines.  In Section
\ref{sec:Comparisons} we compare our results to those recently
published by Brinchmann \etal (2004). Stellar population and emission
line properties are used in Section \ref{sec:Correlations} to
investigate whether the synthesis produces astrophysically plausible
results. Finally, Section \ref{sec:Conclusions} summarizes our main
findings.

\section{Spectral Synthesis}

\label{sec:Synthesis}

\subsection{Method}

\label{sec:SynthesisMethod}

Our synthesis code, which we call STARLIGHT, was first discussed in
CF04 (see also Cid Fernandes \etal 2004b and Garcia-Rissman \etal 2005
for different applications of the same code). STARLIGHT mixes
computational techniques originally developed for empirical population
synthesis with ingredients from evolutionary synthesis
models. Briefly, we fit an observed spectrum $O_\lambda$ with a
combination of $N_\star$ Simple Stellar Populations (SSP) from the
evolutionary synthesis models of BC03.  Extinction is modeled as due
to foreground dust, and parametrized by the V-band extinction
$A_V$. The Galactic extinction law of Cardelli, Clayton \& Mathis
(1989) with $R_V =3.1$ is adopted. Line of sight stellar motions are
modeled by a Gaussian distribution $G$ centered at velocity $v_\star$
and with dispersion $\sigma_\star$.  With these assumptions the model
spectrum is given by

\begin{equation}
\label{eq:model}
M_\lambda = M_{\lambda_0}
    \left[
    \sum_{j=1}^{N_\star} x_j b_{j,\lambda} r_\lambda
    \right]
    \otimes G(v_\star,\sigma_\star)
\end{equation}

\begin{figure*}
\resizebox{\textwidth}{!}{\includegraphics{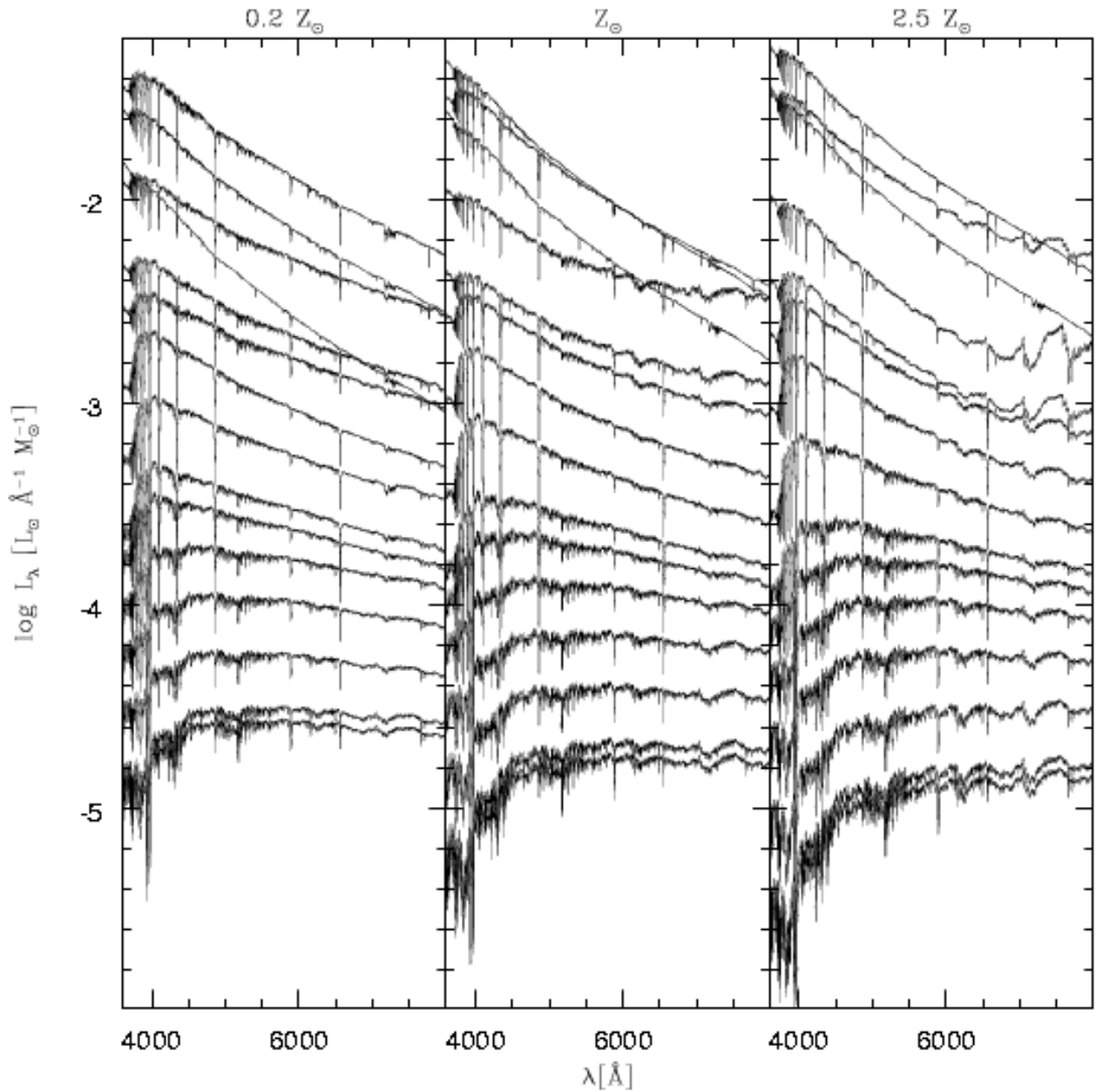}}
\caption{Spectra of the 45 SSPs used in the spectral synthesis (from
BC03). The base comprises 3 different metallicities, $Z = 0.2$, 1 and
2.5 $Z_\odot$, and 15 ages: From top to bottom, $t = $ 0.001, 0.00316,
0.00501, 0.01, 0.02512, 0.04, 0.10152, 0.28612, 0.64054, 0.90479,
1.434, 2.5, 5, 11 and 13 Gyr. All SSPs are normalized to 1 M$_\odot$
at $t = 0$.  }
\label{fig:Base}
\end{figure*}

\ni where $b_{j,\lambda}$ is the spectrum of the $j^{\rm th}$ SSP
normalized at $\lambda_0$, $r_\lambda \equiv 10^{-0.4 (A_\lambda -
A_{\lambda_0})}$ is the reddening term, $M_{\lambda_0}$ is the
synthetic flux at the normalization wavelength, $\vec{x}$ is the {\it
population vector} and $\otimes$ denotes the convolution operator.
Each component $x_j$ ($j = 1\ldots N_\star$) represents the fractional
contribution of the SSP with age $t_j$ and metallicity\footnote{In
this paper we follow the convention used in stellar evolution studies,
which define stellar metallicities in terms of the fraction of mass in
metals. In this system the Sun has $Z_\star = 0.02$.} $Z_j$ to the
model flux at $\lambda_0$. The base components can be equivalently
expressed as a mass fractions vector $\vec{\mu}$.  In this work we
adopt a base with $N_\star = 45$ SSPs, encompassing 15 ages between
$10^6$ and $1.3 \times 10^{10}$ yr and 3 metallicities: $Z = 0.2$, 1
and 2.5 $Z_\odot$. Their spectra, shown in Fig.  \ref{fig:Base}, were
computed with the STELIB library (Le Borgne \etal 2003), Padova 1994
tracks, and Chabrier (2003) IMF (see BC03 for details).

The fit is carried out with a simulated annealing plus Metropolis
scheme which searches for the minimum $\chi^2 = \sum_\lambda \left[
\left(O_\lambda - M_\lambda \right) w_\lambda \right]^2$, where
$w_\lambda^{-1}$ is the error in $O_\lambda$.  Regions around emission
lines, bad pixels or sky residuals are masked out by setting
$w_\lambda = 0$. Pixels which deviate by more than 3 times the rms
between $O_\lambda$ and an initial estimate of $M_\lambda$ are also
given zero weight.

The minimization consists of a series of $N_M = 6$ likelihood-guided
Metropolis explorations of the parameter space. From each iteration to
the next we increase the $w_\lambda$ weights geometrically (which
corresponds to a decrease in the ``temperature'' in statistical
mechanics terms; e.g., MacKay 2003). The step-size in each parameter
is concomitantly decreased and the number of steps is increased. This
scheme gradually focuses on the most likely region in parameter space,
avoiding (through the logic of the Metropolis algorithm) trapping onto
local minima. After completion, the whole fit is fine tuned repeating
the full loop excluding all $x_j = 0$ components. An important
difference (from the computational point of view) with respect to the
code in CF04 is that we now perform a series expansion of the
$r_\lambda$ extinction factor, which allows a much faster computation
of $\chi^2$. Naturally, there are a number of technical parameters in
this complex fitting algorithm, which we have optimized by means of
extensive simulations (see CF04). At any rate, results reported here
are robust with respect to variations in these technical parameters.

\begin{figure*}
\resizebox{\textwidth}{!}{\includegraphics{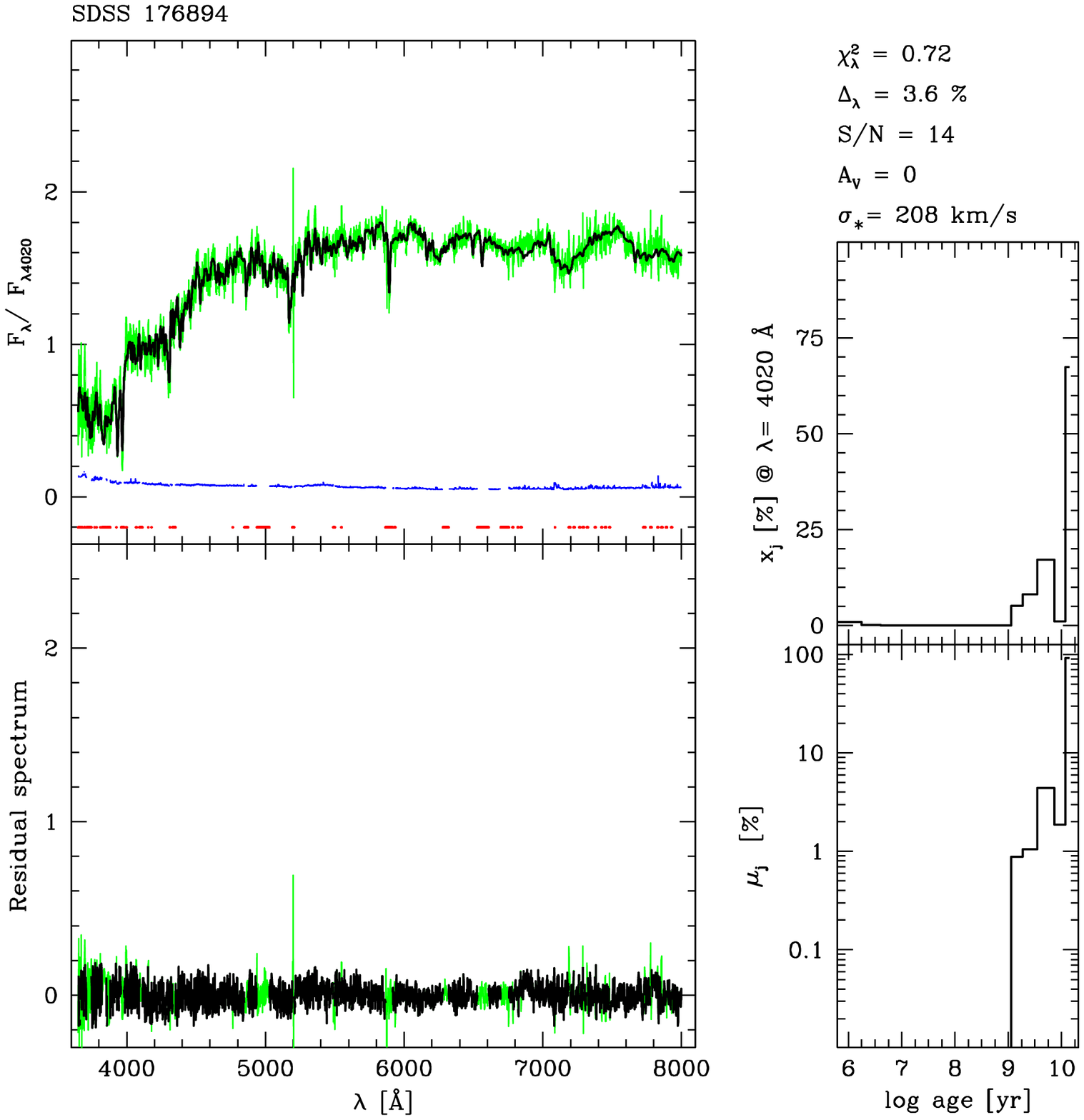}}
\caption{Spectral synthesis of an early type SDSS galaxy. Top-left:
Observed (thin, green line), model (thick, black ) and error spectra
(dashed, blue). Points in the bottom indicate bad pixels (as given by
the SDSS flag) or emission lines windows, both of which were masked
out in the fits. Bottom-left: Residual spectrum. Masked regions are
plotted with a thin, green line. Right: Flux (top) and mass (bottom)
fractions as a function of age.  Some of the derived properties are
listed in the top-right.  $\chi^2_\lambda$ is the reduced $\chi^2$,
and $\Delta_\lambda$ is the mean relative difference between model and
observed spectra; $S/N$ refers to the region around $\lambda_0 = 4020$
\AA.}
\label{fig:examples_fits_1}
\end{figure*}

\begin{figure*}
\resizebox{\textwidth}{!}{\includegraphics{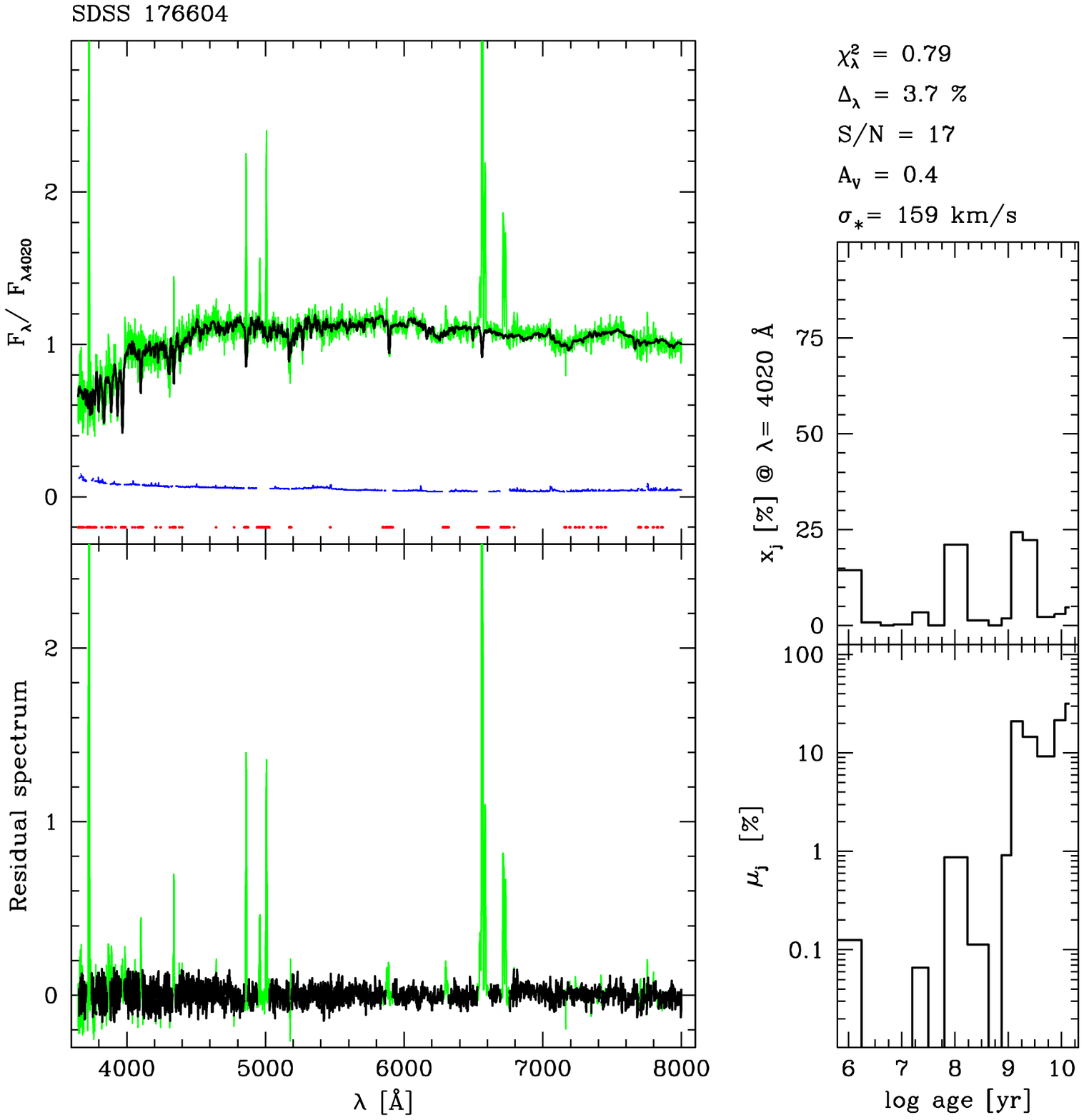}}
\caption{As Fig.~\ref{fig:examples_fits_1} but for a late-type galaxy.}
\label{fig:examples_fits_2}
\end{figure*}

Figs.~\ref{fig:examples_fits_1} and \ref{fig:examples_fits_2}
illustrate the spectral fits obtained for two galaxies drawn from the
SDSS database. The top-left panel shows the observed spectrum (thin
line) and the model (thick), as well as the error spectrum (dashed).
The bottom-left panel shows the $O_\lambda - M_\lambda$ residual
spectrum, while the panels in the right summarize the derived
star-formation history encoded in the age-binned population vector.
These examples, along with those in K03 and CF04, demonstrate that
this simple method is capable of reproducing real galaxy spectra to an
excellent degree of accuracy.

An important application of the synthesis is to measure emission
lines from the residual spectrum, as done by K03 (see also
Section \ref{sec:EmissionLines}). Another, of course, is to infer stellar
population properties from the fit parameters. This was not the
approach followed by K03 and their subsequent papers, who prefer to
derive stellar population properties mainly from the $D_n(4000)$
versus H$\delta_A$ diagram. Our central goal here is to investigate
whether spectral synthesis can also recover reliable stellar
population properties. In the remainder of this section we address
this issue by means of simulations.

\subsection{Robust description of the synthesis results}

\label{sec:RobustDecription}

The existence of multiple solutions is an old known problem in stellar
population synthesis. This multiplicity arises from a combination of
three factors: (1) algebraic degeneracy (number of unknowns larger
than number of observables), (2) intrinsic degeneracies of stellar
populations and (3) measurement uncertainties. Unlike in methods which
synthesize only a handful of spectral indices, algebraic degeneracy
is not a problem for the method outlined above, as the number of
$O_\lambda$ points in any decently sampled spectrum far exceeds the
number of parameters even for large bases. Similarly, by modeling the
whole spectrum one should be able to alleviate degeneracies in
spectral indices associated with different stellar populations
(Jimenez \etal 2004). Measurement errors, however, are still a
problem. A corollary of these complications is that even superb
spectral fits as those shown in Figs. \ref{fig:examples_fits_1} and
\ref{fig:examples_fits_2} do not guarantee that the resulting
parameter estimates are trustworthy.

This brief discussion illustrates the need to assess the degree to
which one can trust the parameters involved in the fit before using
them to infer stellar populations properties. As posed above, the
spectral fits involve $N_\star + 3$ parameters: $N_\star - 1$ of the
$\vec{x}$ components (one degree of freedom is removed by the
normalization constraint), $M_{\lambda_0}$, $A_V$ and the two
kinematical parameters, $v_\star$ and $\sigma_\star$. The reliability
of parameter estimation is best studied by means of simulations which
feed the code with spectra generated with known parameters, add noise,
and then examine the correspondence between input and output values.

CF04 performed this kind of simulation for a $N_\star = 20$ base and
spectra in the 3500--5200 \AA\ interval. Their analysis concentrated
on how well the method recovers $\vec{x}_{\rm input}$ for varying
degrees of noise. The main results of that study are: (1) In the
absence of noise, the method recovers all components of $\vec{x}$ to a
high degree of accuracy.  (2) In the presence of noise, however, the
individual output $x_j$ fractions may deviate drastically from the
input values. Essentially, what happens is that noise washes away the
differences between spectrally similar components, so it becomes
impossible to distinguish them, and the code splits $x_j$ among
neighboring components in spectral space.  This is a common problem
in population synthesis (e.g., Panter \etal 2004; Tremonti 2003; Cid
Fernandes \etal 2001).

Once we have identified the origin of the problem, the remedy to fix
it is evident: bin-over spectrally similar components. In other words,
instead of attempting a fine graded description of stellar populations
mixtures in terms of many ages and metallicities, it is much better to
``marginalize over the details'' and work with a coarser but more
robust description based on combined $x_j$ fractions. As shown by
CF04, condensed versions of the population vector which project its
$N_\star$ components onto just a few physically interesting axes,
yield very reliable results.

Our approach and objectives are conceptually similar to those of the
MOPED code of Heavens, Jimenez \& Lahav (2000; see also Reichardt
\etal 2001, Panter \etal 2004). Operationally, however, the two codes
differ. MOPED compresses data on input by degrading the SDSS spectral
resolution to 20 \AA\ and performing weighted linear combinations of
$O_\lambda$, reducing its dimension to one datum per output parameter
while at the same time minimizing the loss of information. STARLIGHT,
on the other hand, works with the spectrum at its full resolution and
an overdimensioned set of parameters (i.e., large $N_\star$), which we
compress on the description of the output.  The disadvantage of our
approach is computational. Indeed, it is described as a ``slow'' and
``brute force'' method by Panter \etal (2004). This, however, is not a
severe problem given the abundance of CPUs nowadays (STARLIGHT takes
about 4 minutes per galaxy on a 2 GHz Linux-workstation).
Furthermore, there is plenty of room for improvement in the efficiency
of the algorithm, for instance, by implementing a smarter exploration
of the parameter space (Slosar \& Hobson 2003), to the point that
computational constraints could soon become a minor concern.  An
advantage of STARLIGHT is that it also measures kinematical parameters
(mainly $\sigma_\star$) and provides a high resolution template
spectrum.

\subsubsection{Simulations}

\label{sec:Simulations}

We have carried out new simulations designed to test the method and
investigate which combinations of the parameters provide robust
results. These simulations differ from those in CF04 in three main
aspects: (1) the spectral range is now 3650--8000 \AA; (2) the new
base is larger; (3) a more realistic error spectrum was used.  A
further difference is that we use a larger number of iterations, 
partly because we have more base elements than CF04 and partly because
the code is now over 200 times faster.  At each step in the annealing
scheme, the number of steps per parameter is set to $2 L / \epsilon$,
where $L$ is the length of the allowed range for the parameter (e.g.,
$L = 1$ for $x_j$ fractions, which range from 0 to 1), and $\epsilon$
is the step-size (see CF04 for a more detailed description of these
technical aspects). This insures that each parameter can in principle
random-walk twice its whole allowed range, which is an adequate
criterion for Metropolis sampling (e.g., MacKay 2003). Furthermore,
there are 12 stages in the annealing schedule. Overall, over $10^7$
combinations of parameters are sampled for each galaxy.

Several sets of simulations were performed. Given our interest in
modeling SDSS galaxies, here we focus on simulations tailored to match
the characteristics of this data set. Test galaxies were built from
the average $\vec{x}$, $A_V$ and $\sigma_\star$ within 65 boxes in the
mean stellar age versus mean stellar metallicity plane obtained for
the sample described in Section \ref{sec:Syntesis_SDSS}. Each
spectrum covers the 3650--8000 \AA\ range in steps of 1 \AA, the same
sampling of the BC03 models.  We generate 20 perturbed versions of
each synthetic spectrum for each of 5 levels of noise: $S/N = 5$, 10,
15, 20 and 30, where $S/N$ is the signal-to-noise ratio per \AA\
in the region around $\lambda_0 = 4020$ \AA. Unlike in CF04, who used
a flat error spectrum, the error at each $\lambda$ was assumed to
follow a Gaussian distribution with amplitude obtained by scaling a
normalized mean SDSS error spectrum to yield the desired $S/N$ at
$\lambda_0$. This error spectrum decreases by a factor of $\sim 3$
from 3650 to 6200 \AA\ and then increases again by $\sim 1.5$ towards
8000 \AA.  Finally, as for the actual data fits, we mask points around
\oii$\lambda\lambda$3726,3729, \neiii$\lambda3869$, H$\epsilon$,
H$\delta$, H$\gamma$, H$\beta$, \oiii$\lambda\lambda$4959,5007,
\hei$\lambda5876$, NaD$\lambda5890$, \oi$\lambda$6300,
\nii$\lambda$$\lambda$6548,6583, H$\alpha$,
\sii$\lambda\lambda$6717,6731.

These new simulations confirm that the individual components of
$\vec{x}$ are very uncertain, so we skip a detailed comparison between
$\vec{x}_{\rm input}$ and $\vec{x}_{\rm output}$ and jump straight to
results based on more robust descriptions of the synthesis output.

\subsubsection{Condensed population vector}

\begin{figure}
\resizebox{13cm}{!}{\includegraphics{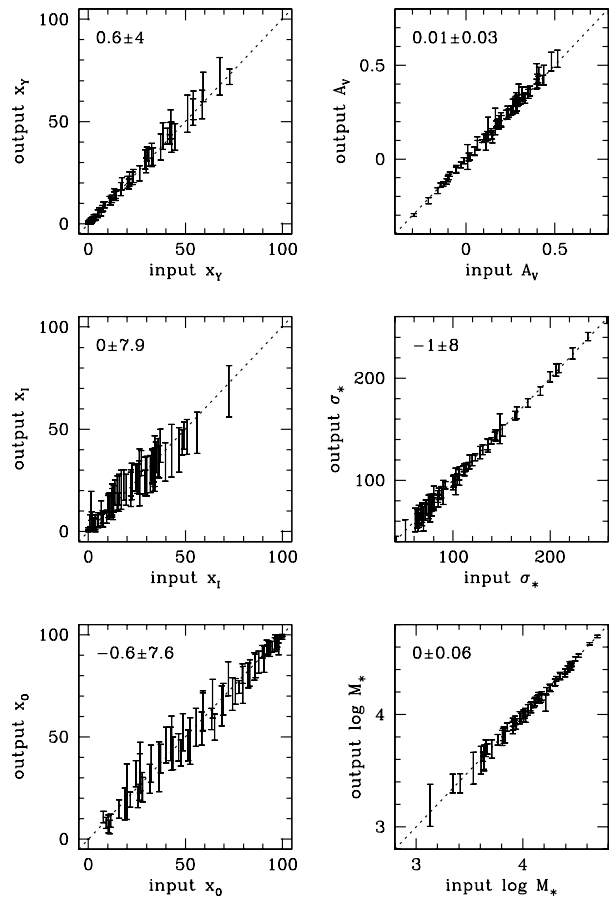}}
\caption{Input versus output synthesis parameters for simulations with
$S/N = 15$ at $\lambda_0 = 4020$ \AA. The units are percentages for
the condensed population vector ($x_Y,x_I,x_O$), magnitudes for $A_V$,
km$\,$s$^{-1}$ for $\sigma_\star$ and arbitrary for $M_\star$.  The
output parameters are represented by $\pm 1$ sigma error bars centered
on the mean values obtained by fitting 20 realizations of each of 65
test galaxies.}

\label{fig:input_x_ouput_1}
\end{figure}

A coarse but robust description of the star-formation history of a
galaxy may be obtained by binning $\vec{x}$ onto ``Young'' ($t_j <
10^8$ yr), ``Intermediate-age'' ($10^8 \le t_j \le 10^9$ yr), and
``Old'' ($t_j > 10^9$ yr) components ($x_Y$, $x_I$ and $x_O$
respectively).  These age-ranges were defined on the basis of the
simulations, by seeking which combinations of $x_j$'s produce smaller
input $-$ output residuals.  Table \ref{tab:Parameter_Uncertainties}
and figure \ref{fig:input_x_ouput_1} show that these 3 components are
very well recovered by the method, with uncertainties smaller than
$\Delta x_Y = 0.05$, $\Delta x_I = 0.1$, and $\Delta x_O = 0.1$ for
$S/N \ge 10$.

\begin{table*}
\begin{centering}
\begin{tabular}{lrrrrr}
\multicolumn{6}{c}{Summary of parameter uncertainties}\\ \hline
   Parameter     &
\multicolumn{5}{c}{$S/N$ at $\lambda = 4020$ \AA} \\
   &
 5 &
10 &
15 &
20 &
30 \\ \hline
$x_Y$          		&  0.35 $\pm$   7.06 &   0.21 $\pm$   4.50 &   0.62 $\pm$   4.04 &   0.56 $\pm$   3.04 &   0.64 $\pm$   2.63 \\ 
$x_I$          		&  0.76 $\pm$  13.94 &  -0.06 $\pm$   9.00 &   0.01 $\pm$   7.88 &  -0.22 $\pm$   6.26 &  -0.02 $\pm$   5.07 \\ 
$x_O$          		& -1.12 $\pm$  13.02 &  -0.15 $\pm$   8.83 &  -0.63 $\pm$   7.61 &  -0.34 $\pm$   6.19 &  -0.62 $\pm$   5.05 \\ 
$\mu_Y$        		&  0.02 $\pm$   1.66 &   0.11 $\pm$   1.41 &   0.16 $\pm$   1.18 &   0.18 $\pm$   1.05 &   0.19 $\pm$   0.80 \\ 
$\mu_I$        		&  1.11 $\pm$  10.26 &   0.99 $\pm$   7.57 &   0.93 $\pm$   6.10 &   1.00 $\pm$   5.03 &   1.17 $\pm$   4.22 \\ 
$\mu_O$        		& -1.13 $\pm$  10.54 &  -1.10 $\pm$   8.17 &  -1.09 $\pm$   6.54 &  -1.18 $\pm$   5.59 &  -1.36 $\pm$   4.63 \\ 
$A_V$          		&  0.01 $\pm$   0.09 &   0.00 $\pm$   0.05 &   0.01 $\pm$   0.03 &   0.01 $\pm$   0.03 &   0.00 $\pm$   0.02 \\ 
$v_\star$      		&  0.24 $\pm$  17.73 &  -0.11 $\pm$   8.55 &   0.07 $\pm$   5.92 &   0.00 $\pm$   4.23 &   0.05 $\pm$   2.81 \\ 
$\sigma_\star$ 		& -3.12 $\pm$  24.32 &  -1.40 $\pm$  12.36 &  -1.01 $\pm$   7.71 &  -0.75 $\pm$   5.73 &  -0.57 $\pm$   3.78 \\ 
$\log M_\star$ 		&  0.01 $\pm$   0.11 &  -0.01 $\pm$   0.08 &  -0.01 $\pm$   0.06 &  -0.01 $\pm$   0.05 &  -0.02 $\pm$   0.04 \\ 
$<\log t_\star>_L$ 	& -0.01 $\pm$   0.14 &   0.01 $\pm$   0.08 &   0.00 $\pm$   0.06 &   0.01 $\pm$   0.05 &   0.01 $\pm$   0.04 \\ 
$<\log t_\star>_M$	& -0.03 $\pm$   0.20 &  -0.04 $\pm$   0.14 &  -0.03 $\pm$   0.11 &  -0.04 $\pm$   0.10 &  -0.04 $\pm$   0.08 \\ 
$\log <Z_\star>_L$	& -0.01 $\pm$   0.15 &  -0.01 $\pm$   0.09 &   0.00 $\pm$   0.08 &   0.00 $\pm$   0.06 &   0.00 $\pm$   0.05 \\ 
$\log <Z_\star>_M$      & -0.03 $\pm$   0.18 &  -0.02 $\pm$   0.13 &  -0.03 $\pm$   0.11 &  -0.02 $\pm$   0.09 &  -0.03 $\pm$   0.08 \\ 
$\sigma_L(\log t_\star)$&  0.03 $\pm$   0.16 &   0.00 $\pm$   0.10 &   0.00 $\pm$   0.08 &  -0.01 $\pm$   0.07 &  -0.01 $\pm$   0.06 \\ 
$\sigma_M(\log t_\star)$& -0.04 $\pm$   0.09 &  -0.02 $\pm$   0.06 &  -0.01 $\pm$   0.05 &  -0.01 $\pm$   0.04 &   0.00 $\pm$   0.04 \\ 
$\sigma_L(Z_\star)$	& -0.01 $\pm$   0.21 &   0.00 $\pm$   0.14 &   0.02 $\pm$   0.11 &   0.02 $\pm$   0.10 &   0.02 $\pm$   0.09 \\ 
$\sigma_M(Z_\star)$	& -0.11 $\pm$   0.26 &  -0.05 $\pm$   0.19 &  -0.03 $\pm$   0.16 &  -0.02 $\pm$   0.14 &  -0.02 $\pm$   0.12 \\ 
$\chi^2/N_\lambda$      &  0.95 $\pm$   0.02 &   0.95 $\pm$   0.02 &   0.95 $\pm$   0.02 &   0.95 $\pm$   0.02 &   0.95 $\pm$   0.02 \\ 
$\Delta_\lambda$        & 11.64 $\pm$   4.05 &   5.37 $\pm$   1.19 &   3.54 $\pm$   0.78 &   2.65 $\pm$   0.58 &   1.76 $\pm$   0.39 \\ 
$\Delta_\lambda \times (S/N)_{\lambda_0}$ 
                        &  0.78 $\pm$   0.27 &   0.68 $\pm$   0.20 &   0.64 $\pm$   0.21 &   0.59 $\pm$   0.22 &   0.51 $\pm$   0.23 \\ 
\hline
\end{tabular}
\end{centering}
\caption{Each line corresponds to a parameter constructed by combining
the parameters given by the synthesis. The different columns list the
mean $\pm$ rms difference between output and input values of the
corresponding quantity, as obtained from simulations with different
signal-to-noise ratios. For instance, the mean output minus input
difference in $<\log t_\star>_L$ for $S/N = 10$ is 0.01 dex and the
rms dispersion is 0.08 dex. The last 3 lines describe the statistics
of goodness-of-fit indicators. The units are percentage for light and
mass fractions ($x_Y \ldots \mu_O$), mag for $A_V$, km$\,$s$^{-1}$ for
$v_\star$ and $\sigma_\star$, percentage for $\Delta_\lambda$ and dex
for logarithmic quantities.}
\label{tab:Parameter_Uncertainties}
\end{table*}

\subsubsection{Mass, extinction and velocity dispersion}

\label{sec:Mass_and_vd}

Fig. \ref{fig:input_x_ouput_1} also shows the input versus output
values of $A_V$, $\sigma_\star$ and the stellar mass $M_\star$. The
latter is not an explicit input parameter of the models, but may be
computed from $\vec{\mu}$ and the $M_\star/L_{\lambda_0}$ ratio of the
different populations in the base. The uncertainties in these
parameters are $\Delta A_V < 0.05$ mag, $\Delta \log M_\star < 0.1$
dex and $\Delta \sigma_\star < 12$ km$\,$s$^{-1}$ for $S/N \ge 10$.
(Table \ref{tab:Parameter_Uncertainties}).

\subsubsection{Mean stellar age}

\label{sec:Mean_Stellar_Age}

If one had to choose a single parameter to characterize the stellar
population mixture of a galaxy, the option would certainly be its mean
age. We define two versions of mean stellar age (the logarithm of the age,
actually), one weighted by light

\begin{equation}
\label{eq:mean_age_flux}
<\log t_\star>_L =  \sum_{j = 1}^{N_\star} x_j \log t_j
\end{equation}

\ni and another weighted by stellar mass

\begin{equation}
\label{eq:mean_age_mass}
<\log t_\star>_M =  \sum_{j = 1}^{N_\star} \mu_j \log t_j
\end{equation}

\ni Note that, by construction, both definitions are limited to the 1
Myr--13 Gyr range spanned by the base. The mass weighted mean age is
in principle more physical, but, because of the non-constant $M/L$ of
stars, it has a much less direct relation with the observed spectrum
than $<\log t_\star>_L$.

\begin{figure*}
\resizebox{13cm}{!}{\includegraphics{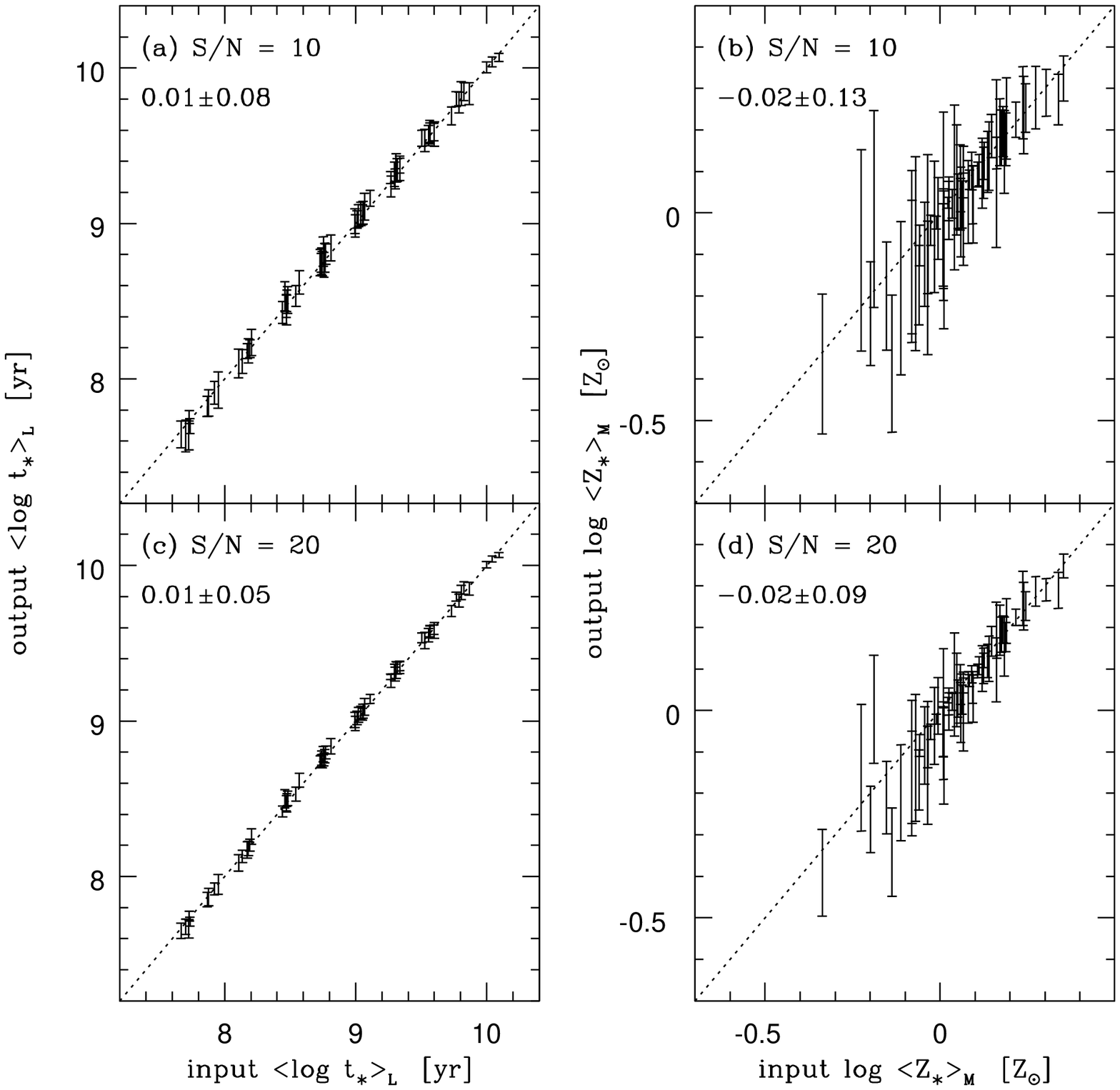}}
\caption{Input versus output mean stellar ages and metallicities for
simulations with $S/N = 10$ and 20.}
\label{fig:input_x_ouput_mean_age_and_Z}
\end{figure*}

Fig. \ref{fig:input_x_ouput_mean_age_and_Z} shows the input against
output $<\log t_\star>_L$ for simulations with $S/N = 10$ and 20. The
plots show that the mean age is a very robust quantity. The rms
difference between input and output $<\log t_\star>_L$ values is $\le
0.08$ dex for $S/N > 10$, and $\le 0.14$ dex for $<\log t_\star>_M$
(Table \ref{tab:Parameter_Uncertainties}). Although the uncertainties
of $<\log t_\star>_L$ and $<\log t_\star>_M$ are comparable in
absolute terms, the latter index spans a smaller dynamical range
(because of the large $M/L$ ratio of old populations), so in practice
$<\log t_\star>_L$ is the more useful of the two indices.

Given that the mean stellar age is so well recovered by the method one
might attempt more detailed descriptions of the star-formation history
involving, say, higher moments of the age distribution. For instance,

\begin{equation}
\label{eq:rms_age_flux}
\sigma_L(\log t_\star) =  \left[\sum_{j = 1}^{N_\star} x_j
\left(\log t_j - <\log t_\star>_L\right)^2\right]^{1/2}
\end{equation}

\ni measures the flux-weighted standard deviation of the log age
distribution, and might be useful to distinguish galaxies dominated by
a single population from those which had continuous or bursty
star-formation histories. The uncertainty in this index is of order
0.1 dex.

\subsubsection{Mean stellar metallicity}

\label{sec:Mean_Stellar_Z}

Given an option of what to chose as a second parameter to describe a
mixed stellar population, the choice would likely be its typical
metallicity.  Analogously to what we did for ages, we define both
light and mass-weighted mean stellar metallicities:

\begin{equation}
\label{eq:mean_Z_flux}
<Z_\star>_L =  \sum_{j = 1}^{N_\star} x_j Z_j
\end{equation}

\ni and

\begin{equation}
\label{eq:mean_Z_mass}
<Z_\star>_M =  \sum_{j = 1}^{N_\star} \mu_j Z_j
\end{equation}

\ni both of which are bounded by the 0.2--2.5 $Z_\odot$ base limits.
Fig. \ref{fig:input_x_ouput_mean_age_and_Z} and Table
\ref{tab:Parameter_Uncertainties} show that the rms of $\Delta \log
<Z_\star>_M = \log <Z_\star>_{M,\rm output} - \log <Z_\star>_{M,\rm
input}$ is of order 0.1 dex. In absolute terms this is comparable to
$\Delta <\log t_\star>$, but note that $<Z_\star>$ covers a much
narrower dynamical range than $<\log t_\star>$, so that in practice
mean stellar metallicities are more sensitive to errors than mean
ages. This is not surprising, given that age is the main driver of
variance among SSP spectra, metallicity having a ``second-order''
effect (e.g., Schmidt \etal 1991; Ronen, Aragon-Salamanca \& Lahav
1999). This is the reason why studies of the stellar populations of
galaxies have a much harder time estimating metallicities than ages,
to the point that one is often forced to bin-over the $Z$ information
and deal only with age-related estimates such as $<\log t_\star>$ (e.g.,
Cid Fernandes \etal 2001; Cid Fernandes, Le\~ao \& Rodrigues Lacerda
2003; K03).

Notwithstanding these notes, it is clear that uncertainties of $\sim
0.1$ dex in $<Z_\star>$ are actually good news, since they do allow us
to recover useful information on an important but hard to measure
property. This new tracer of stellar metallicity is best applicable to
large samples of galaxies such as the SDSS. The statistics of samples
help reducing uncertainties associated with $<Z_\star>$ estimates for
single objects and allows one to investigate correlations between
$<Z_\star>$ and other galaxy properties (Sodr\'e \etal, in
preparation; Section \ref{sec:Stellar_X_Nebular_Metallicities}).

\subsubsection{Age-Metallicity degeneracy}

\begin{figure*}
\psfig{file=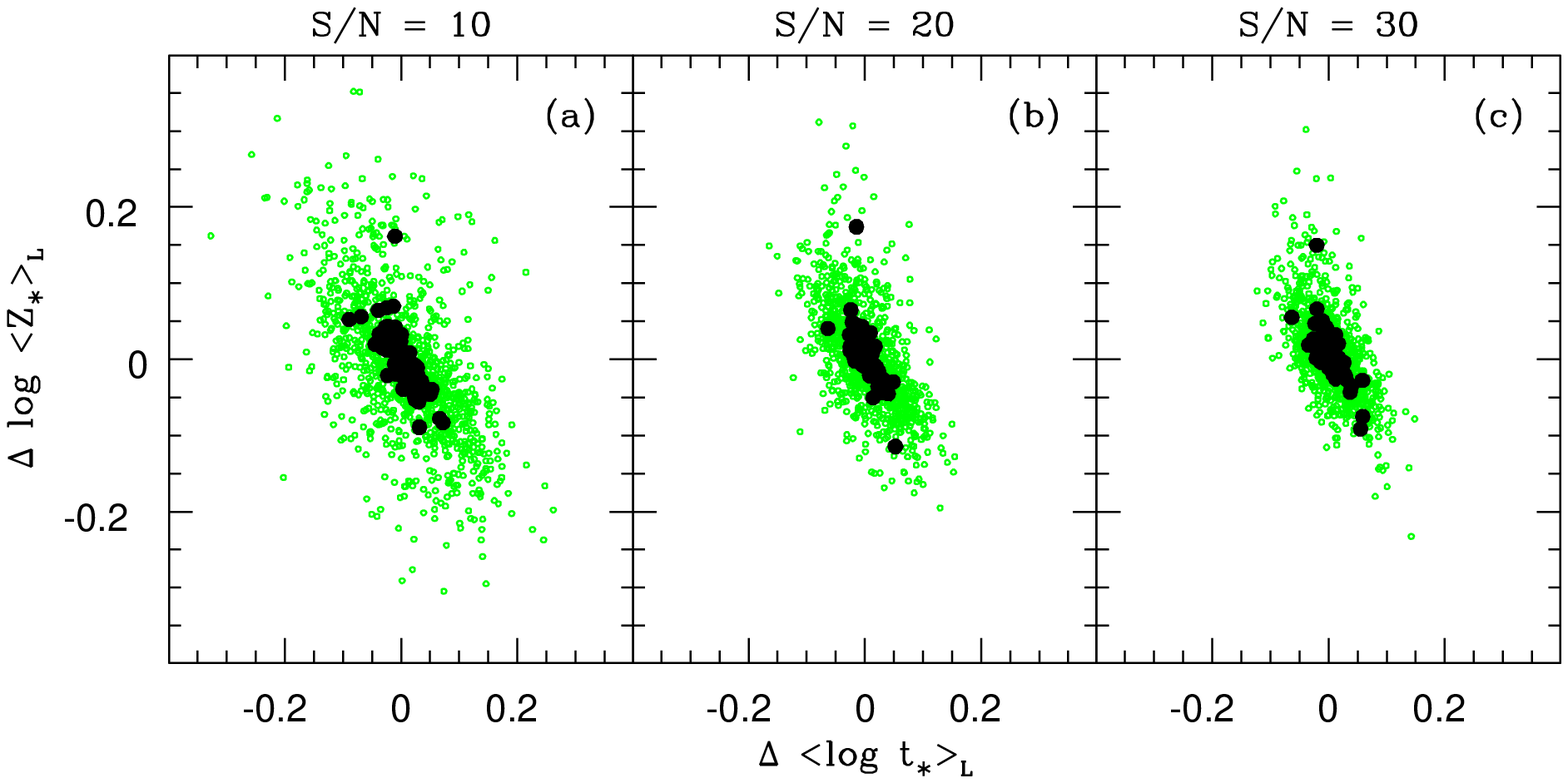,width=18cm,bbllx=20,bblly=430,bburx=600,bbury=710}
\caption{Output minus input residuals in $\log <Z_\star>_L$ against
residuals in $<\log t_\star>_L$. Small circles are the individual
test galaxies. Large filled circles mark the mean residuals for 20
perturbed versions of the same test galaxy. The anti-correlation
illustrates the effect of the age-metallicity degeneracy.}
\label{fig:age_Z_degeneracy}
\end{figure*}

Our method tends to underestimate $<Z_\star>$ for metal-rich systems
and vice-versa. This bias is due to the infamous age-metallicity
degeneracy.  In order to verify to which degree our synthesis is
affected by this well known problem (e.g., Renzini \& Buzzoni 1986;
Worthey 1994; Bressan, Chiosi \& Tantalo 1996) we have examined the
correlation between the output minus input residuals in $<\log
t_\star>_L$ and $\log <Z_\star>_L$. The age-$Z$ degeneracy acts in the
sense of confusing old, metal-poor systems with young, metal-rich ones
and vice-versa, which should produce anti-correlated residuals. These
are indeed seen in figure \ref{fig:age_Z_degeneracy}, where small
symbols represent all 1300 individual simulations and filled circles
correspond to the mean residuals obtained for each set of 20
perturbations of the 65 test galaxies. This anti-correlation is also
present with the mass-weighted mean stellar metallicity $<Z_\star>_M$,
but is not as strong as for $<Z_\star>_L$. On the other hand, the
uncertainty in $<Z_\star>_M$ is always larger than for $<Z_\star>_L$
(Table \ref{tab:Parameter_Uncertainties}).

The age-$Z$ degeneracy is thus present in our method, introducing
systematic biases in our $<Z_\star>$ and $<\log t_\star>$ estimates at
the level of up to $\sim 0.1$--0.2 dex. None of the results reported
in this paper rely on this level of precision.

\subsection{Simulations for random parameters}

We have also carried out a separate set of simulations for 100 test
galaxies generated by random combinations of $\vec{x}$, $A_V$,
$\sigma_\star$. These simulations differ from the ones presented above
in that the test galaxies are not restricted to match the range of
properties inferred from the application of the synthesis to SDSS
galaxies (Section \ref{sec:Syntesis_SDSS}). Given that these random
galaxies span a broader (but less representative of our sample
galaxies) region of the parameter space, one might expect to find
larger uncertainties in the derived properties.  This was indeed
confirmed in the numerical experiments, although the effect is small.
The uncertainties in the $(x_Y,x_I,x_O)$ binned fractions increase by
just a couple of percentage points with respect to those in Table
\ref{tab:Parameter_Uncertainties}. Similarly, the increase in $\Delta
A_V$ is of just 0.02 mag, while $\Delta <\log t_\star>_L$ and $\Delta
\log <Z_\star>_M$ increase by $\sim 0.02$ and 0.03 dex respectively.

The only properties whose uncertainties are substantially larger that
those reported in Table \ref{tab:Parameter_Uncertainties} are the
stellar mass and velocity dispersion. For instance, while the
SDSS-based simulations for $S/N = 10$ yield $\Delta \log M_\star =
0.08$ dex and $\Delta \sigma_\star = 12$ km$\,$s$^{-1}$, with random
galaxies these errors increase to 0.15 dex and 24 km$\,$s$^{-1}$
respectively.  The reason for this apparent discrepancy is due to the
fact that the set of random test-galaxies contains a larger proportion
of systems dominated by very young stellar populations than we find
for SDSS galaxies. An error $\Delta x_Y$ in the light fraction
associated to these populations must be compensated by errors in the
older components, which carry most of the mass, even when $x_Y$ is
large. As illustrated by the size of the error-bars in figure
\ref{fig:input_x_ouput_1}, $\Delta x_Y$ increases as $x_Y$ increases,
so the errors in the mass fractions components increase too, leading
to larger dispersion in $M_\star$. Furthermore, galaxies with large
$x_Y$ have few absorption features to constrain the kinematical
broadening of the spectrum, which explains the larger dispersion in
our $\sigma_\star$ estimates. To prove this point, we have
re-evaluated the uncertainties in $M_\star$ and $\sigma_\star$
excluding test galaxies with $x_Y > 70$\%, which corresponds to
systems which formed $\ga 20$\% of their stellar mass over the past $<
10^8$ yr.  For this subset of the simulations, which comprise 72
galaxies (each one split onto 20 different spectra corresponding to
independent Monte Carlo realizations of the noise), the uncertainties
in mass and velocity dispersion decrease to $\Delta \log M_\star =
0.09$ dex and $\Delta \sigma_\star = 14$ km$\,$s$^{-1}$ (for $S/N =
10$), only slightly larger than those reported in Section
\ref{sec:Mass_and_vd}.  Uncertainties in other properties also
decrease to values very similar to those listed in Table
\ref{tab:Parameter_Uncertainties}.

We thus conclude that the parameter uncertainties studied in Section
\ref{sec:RobustDecription} and summarized in Table
\ref{tab:Parameter_Uncertainties} are only moderately affected by the
design of the simulations, and represent fair estimates of the
limitations of our synthesis method.

Finally, we have carried out simulations using the $Z = 0.02 Z_\odot$
BC03 SSPs to generate further test galaxies. Galaxies with such low
metallicity are not expected to be present in significant numbers in
the sample described in Section \ref{sec:Sample_definition}, given
that it excludes low luminosity systems like HII galaxies and dwarf
ellipticals (which are also the least metallic ones by virtue of the
mass-metallicity relation).  Still, it is interesting to investigate
what would happen in this case. When synthesized with our 0.2--$2.5
Z_\odot$ base (Fig.~\ref{fig:Base}), these extremely metal poor
galaxies are modeled predominantly with the $0.2 Z_\odot$ components,
as intuitively expected.  Moreover, the mismatch in metallicity
introduces non-negligible biases in other properties, like masses,
mean ages and extinction ($M_\star$, for instance, is systematically
underestimated by 0.3 dex). Similar problems should be encountered
when modeling systems with $Z > 2.5 Z_\odot$. These results serve as
a reminder that our base spans a wide but finite range in stellar
metallicity, and that extrapolating these limits has an impact on the
derived physical properties. While there is no straightforward {\it a
priori} diagnostic of which galaxies violate these limits, in general,
one should be suspicious of objects with mean $Z_\star$ too close to
the base limits.

\subsection{Summary of the simulations}

Summarizing this theoretical study, we have performed simulations
designed to evaluate the accuracy of our spectral synthesis method.
The simulations mimic in as much as possible the wavelength range,
spectral resolution, error spectrum, and $S/N$ of the actual SDSS data
studied below. Several physically motivated combinations of the
synthesis parameters were investigated to establish their precision at
different $S/N$. Table \ref{tab:Parameter_Uncertainties} summarizes
the uncertainties in these quantities and a few additional ones not
explicitly mentioned above.  In what follows we focus on 5 parameters:
Stellar mass, velocity dispersion, extinction, mean stellar ages and
metallicities, all of which were found to be adequately recovered.

This exercise demonstrates that we are capable of producing reliable
estimates of several parameters of astrophysical interest, at least in
principle.  We must nevertheless emphasize that this conclusion relies
entirely on models and on an admittedly simplistic view of
galaxies. When applying the synthesis to real galaxy spectra, a series
of other effects come into play. For instance, the extinction law
appropriate for each galaxy likely differs from the one used here: in
our Galaxy, the ratio of total to selective extinction of stellar
sources, $R_V = A_V/E(B-V)$, is known to depend on the line of sight
(Cardelli \etal 1989; Patriarchi \etal 2001); in addition, the
wavelength dependence of the attenuation of light from an extended
source such as a galaxy includes the effects of scattering back into
the light beam, and depends on the relative distribution of stars and
dust (Witt \etal 1992; Gordon \etal 1997). Also, one might expect that
each population of stars is affected by a distinct extinction (Panuzzo
\etal 2003; Charlot \& Fall 2000). Similarly, while in the
evolutionary tracks adopted here the metal abundances are scaled from
the solar values, non-solar abundance mixtures are known to occur in
stellar systems (e.g., Trager \etal 2000a,b), not to mention
uncertainties in the SSP models and the always present issue of the
IMF. In short, evidence against these simplifying hypothese abound.

Accounting for all these effects in a consistent way is not currently
feasible. We mention these caveats not to dismiss simple models, but
to highlight that all parameter uncertainties discussed above are
applicable within the scope of the model.  Hence, while the
simulations lend confidence to the synthesis method, one might remain
skeptical of its actual power. The next sections further address the
reliability of the synthesis, this time from a more empirical
perspective.

\section{Analysis of a volume-limited galaxy sample}

\label{sec:Syntesis_SDSS}

In this section we apply our synthesis method to a large sample of
SDSS galaxies to estimate their stellar population properties. We also
present measurements of emission line properties, obtained from the
observed minus synthetic spectra. The information provided by the
synthesis of so many galaxies allows one to address a long menu of
astrophysical issues related to galaxy formation and evolution.
Before venturing in the exploration of such issues, however, it is
important to validate the results of the synthesis by as many means as
possible. Hence, the goal of the study presented below is not so much
to explore the physics of galaxies but to provide an empirical test of
our synthesis method.  The results reported in this section are used
in Sections \ref{sec:Comparisons} and \ref{sec:Correlations} with this
purpose.

\subsection{Sample definition}

\label{sec:Sample_definition}

The spectroscopic data used in this work were taken from the
SDSS. This survey provides spectra of objects in a large wavelength
range (3800--9200 \AA) with mean spectral resolution
$\lambda/\Delta\lambda \sim 1800$, taken with 3 arcsec diameter
fibers. The most relevant characteristic of this survey for our study
is the enormous amount of good quality, homogeneously obtained
spectra. The data analyzed here were extracted from the SDSS main
galaxy sample available in the Data Release 2 (DR2; Abazajian \etal
2004). This flux-limited sample consists of galaxies with
reddening-corrected Petrosian $r$-band magnitudes $r \le 17.77$, and
Petrosian $r$-band half-light surface brightnesses $\mu_{50} \le 24.5$
mag arcsec$^{-2}$ (Strauss \etal 2002).

From the main sample, we first selected spectra with a redshift
confidence $\ge 0.35$. Following the conclusions of Zaritsky,
Zabludoff, \& Willick (1995), we have imposed a redshift limit of $z >
0.05$ (trying to avoid aperture effects and biases; see e.g. G\'omez
\etal 2003) and selected a volume limited sample up to $z = 0.1$,
corresponding to a $r$-band absolute magnitude limit of $M(r)= -20.5$.
The absolute magnitudes used here are k-corrected with the help of the
code provided by Blanton \etal (2003; \texttt{kcorrect v3\_2}) and
assuming the following cosmological parameters: $H_0$ = 70 km s$^{-1}$
Mpc$^{-1}$, $\Omega_M = 0.3$ and $\Omega_\Lambda = 0.7$.  We also
restricted our sample to objects for which the observed spectra show a
$S/N$ ratio in $g$, $r$ and $i$ bands greater than 5.  These
restrictions leave us with a volume limited sample containing $50362$
galaxies, which leads to a completeness level of $\sim$ 98.5 per cent.

\subsection{Results of the spectral synthesis}

\label{sec:SDSS_spectral_fits}

All 50362 spectra were brought to the rest-frame (using the redshifts
in the SDSS database), sampled from 3650 to 8000 \AA\ in steps of 1
\AA, corrected for Galactic extinction\footnote{Unlike in the first
data release, the final calibrated spectra from the DR2 are not
corrected for foreground Galactic reddening.} using the maps given by
Schlegel, Finkbeiner \& Davis (1998) and the extinction law of
Cardelli \etal (1989, with $R_V = 3.1)$, and normalized by the median
flux in the 4010--4060 \AA\ region. The $S/N$ ratio in this spectral
window spans the 5--30 range, with median value of 14. Besides the
masks around the lines listed in Section \ref{sec:Simulations}, we exclude
points with SDSS flag $\ge 2$, which signals bad pixels, sky residuals
and other artifacts. After this pre-processing, the spectra are fed
into the STARLIGHT code described in Section \ref{sec:SynthesisMethod}. On
average, the synthesis is performed with $N_\lambda = 3677$ points,
after discounting the ones which are clipped by our $\le 3$ sigma
threshold (typically 40 points) and the masked ones.

The spectral fits are generally very good, as illustrated in Figs.\
\ref{fig:examples_fits_1} and \ref{fig:examples_fits_2}. The mean
value of $\chi^2/N_\lambda$ is 0.78. In fact, this is somewhat too
good, since from the simulations we expect $\chi^2/N_\lambda \sim
0.95$. This is a minor difference, which could be fixed decreasing the
errors in $O_\lambda$ by $\sim 10$\%. We further quantify the quality
of the fits by $\Delta_\lambda$, the mean value of $|O_\lambda -
M_\lambda|/O_\lambda$ over all non-masked points. From the simulations
we expect this alternative figure of merit to be of order of 0.6 times
the noise-to-signal ratio at $\lambda_0$ (Table
\ref{tab:Parameter_Uncertainties}). This is exactly the mean value of
$\Delta_\lambda \times S/N$ in the actual fits, again indicating
acceptable fits.

The total stellar masses of the galaxies were obtained from the
stellar masses derived from the spectral synthesis (which correspond
to the light entering the fibers) by dividing them by $(1 - f)$,
where $f$ is the fraction of the total galaxy luminosity in the
$z$-band outside the fiber.  This approach, which neglects stellar
population and extinction gradients, leads to an increase of typically
0.5 in $\log M_\star$.  We did not apply any correction to the
velocity dispersion estimated by the code given that the spectral
resolution of the BC03 models and the data are very similar.

We point out that we did not constrain the extinction $A_V$ to be
positive. There are several reasons for this choice: (a) some objects
may be excessively dereddened by Galactic extinction; (b) some objects
may indeed require bluer SSP spectra than those in the base; (c) the
observed light may contain a scattered component, which would induce a
bluening of the spectra not taken into account by the adopted pure
extinction law; (d) constraining $A_V$ to have only positive values
produces an artificial concentration of solutions at $A_V = 0$, an
unpleasant feature in the $A_V$ distribution. Interestingly, most of
the objects for which we derive negative $A_V$ (typically $-0.1$ to
$-0.3$ mag) are early-type galaxies. These galaxies are dominated by
old populations, and expected to contain little dust. This is
consistent with the result of K03, who find negative extinction
primarily in galaxies with a large $D_n(4000)$. The distribution of
$A_V$ for these objects, which can be selected on the basis of
spectral or morphological properties, is strongly peaked around $A_V =
0$, so that objects with $A_V < 0$ can be considered as consistent
with having zero extinction.  In any case, none of the results
reported in this paper is significantly affected by this choice.

\subsection{Emission line measurements}

\label{sec:EmissionLines}

Besides providing estimates of stellar population properties, the
synthesis models allow the measurement of emission lines from the
``pure-emission'', starlight subtracted spectra $(O_\lambda -
M_\lambda)$. We have measured the lines of
\oii$\lambda\lambda$3726,3729, \oiii$\lambda$4363, H$\beta$,
\oiii$\lambda\lambda$4959,5007, \oi$\lambda$6300, \nii$\lambda$6548,
H$\alpha$, \nii$\lambda$6584 and \sii$\lambda\lambda$6717,6731. Each
line was treated as a Gaussian with three parameters: width, offset
(with respect to the rest-frame central wavelength), and flux.  Lines
from the same ion were assumed to have the same width and offset. We
have further imposed \oiii$\lambda5007$/\oiii$\lambda4959 = 2.97$ and
\nii$\lambda6584$/\nii$\lambda6548 = 3$ flux ratio constraints.
Finally, we consider a line to have significant emission if its fit
presents a $S/N$ ratio greater than 3.

In some of the following analysis, galaxies with emission lines are
classified according to their position in the \oiii/H$\beta$ versus
\nii/H$\alpha$ diagram proposed by Baldwin, Phillips \& Terlevich
(1981) to distinguish normal star-forming galaxies from galaxies
containing active galactic nuclei (AGN). We define as normal
star-forming galaxies those galaxies that appear in this diagram and
are below the curve defined by K03 (see also Brinchmann \etal
2004). Objects above this curve are transition objects and galaxies
containing AGN.

\subsection{Aperture bias}

\label{sec:ApertureBias}

A point that deserves mention here is the bias that may be introduced
in the analysis due to the use of small fibers to measure the galaxy
spectra. This effect, known as aperture bias, may introduce a redshift
dependence in the measured galaxy spectra, since the fraction of
galaxy light received by a fiber increases with increasing
distance. Zaritsky \etal (1995), based on an analysis of spectra from
the Las Campanas Redshift Survey, find that a lower limit on redshifts
of $z = 0.05$ minimizes the aperture bias. This effect has also been
discussed by K03, G\'omez \etal (2003) and Tremonti \etal (2004) for
SDSS spectra. For instance, these authors found that the galaxy
$z$-band $M/L$ ratio and the gas-phase oxygen abundance decrease by
$\sim$ 0.1 dex over the redshift range of the survey, indicating that
these quantities are moderately affected by aperture bias. The
redshift range considered in our study is smaller than that of the
entire survey, implying that the aperture effects are expected to be
even smaller in our sample.

\begin{figure*}
\centerline{\includegraphics[scale=0.90]{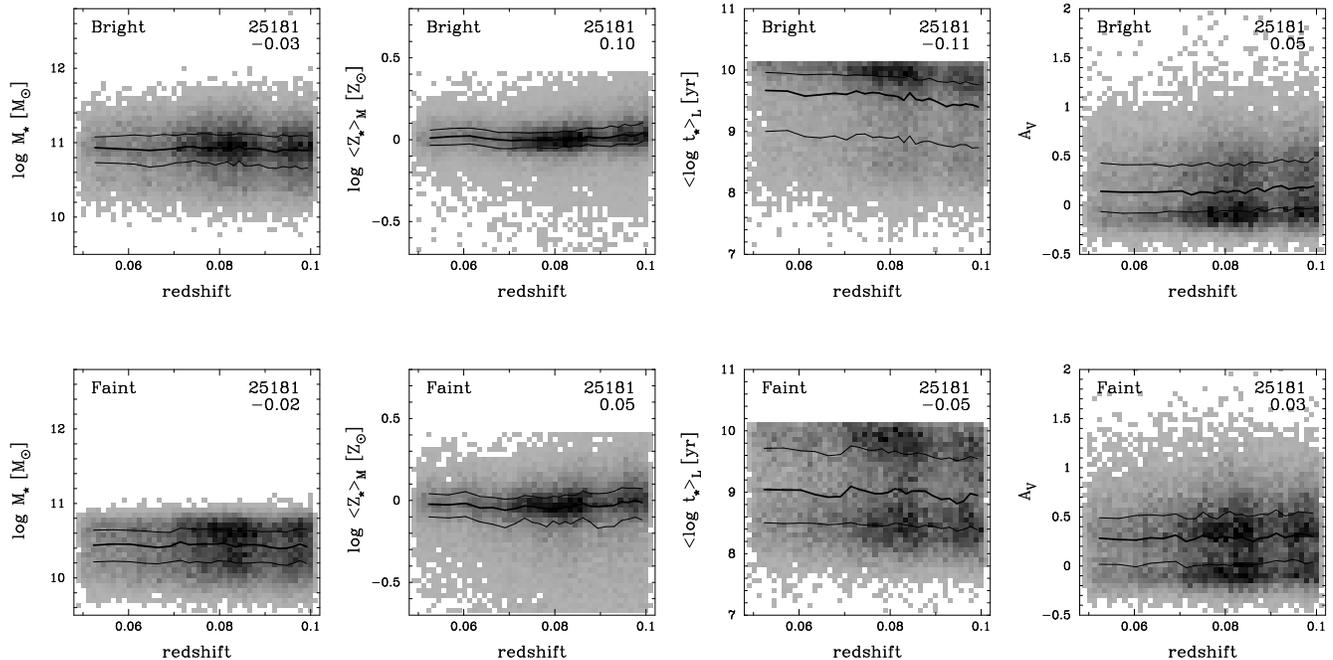}}
\caption{Trends of median values (and respective quartiles) of the
stellar mass, mean stellar metallicity, mean stellar age and the
V-band extinction as a function of redshift. Results are shown for
galaxies with luminosities above (``Bright'', top panels) and below
(``Faint'', bottom panels) the median luminosity of the sample. In
this and following figures the gray scale level represents the number
of galaxies in each pixel, darker pixels having more galaxies; the
upper number in the top right of each panel indicates the total number
of galaxies in the plot, and the lower one is the Spearman rank
correlation coefficient ($r_S$).}
\label{fig:ApertureBias}
\end{figure*}

In order to verify whether our results are significantly affected by
this bias, we investigated the behaviour of some of the quantities
resulting from our analysis ($M_\star$, $<Z_\star>$, $<\log t_\star>$
and $A_V$) as a function of redshift for galaxies with luminosities
above and below $M(r)=-21.17$, the median luminosity of our sample. We
divided the sample in several redshift bins containing the same number
of objects, and computed the median value and the quartiles of each
quantity in each bin. Fig. \ref{fig:ApertureBias} shows as solid lines
the median values of each distribution, as well as their respective
quartiles. None of the quantities seems to be strongly affected by
this bias. The largest correlation with $z$ appears for light-weighted
ages, corresponding to a change along our redshift distribution of
about $0.1$ and $0.26$ dex for faint and bright galaxies,
respectively. This is a plausible result since the light of nearby
objects seen by the fiber aperture is dominated by the older stellar
populations of their bulges. For the other quantities the variations
are well below 0.1 dex.

\section{Comparisons with the MPA/JHU database}

\label{sec:Comparisons}

The SDSS database has been explored by several groups, using different
approaches and techniques.  The MPA/JHU group has recently publicly
released catalogues\footnote{available at
http://www.mpa-garching.mpg.de/SDSS/} of derived physical properties
for 211894 SDSS galaxies, including 33589 narrow-line AGN (K03, see
also Brinchmann \etal 2004). These catalogues are based on the K03
method to infer the star formation histories, dust attenuation and
stellar masses of galaxies from the simultaneous analysis of the 4000
\AA\ break strength, $D_n(4000)$, and the Balmer line absorption index
H$\delta_A$. These two indices are used to constrain the mean stellar
ages of galaxies and the fractional stellar mass formed in bursts over
the past few Gyr, and a comparison with broad-band photometry then
allows to estimate the extinction and stellar masses.

The MPA/JHU catalogues provide very useful benchmarks for similar
studies. In this section we compare the values of some of the
parameters from these catalogues with our own estimates. Catalogues of
galaxy properties obtained with our synthesis method will be made
available in due course. Besides comparing directly measurements of
physical quantities, our aim is also to highlight the differences that
may appear due to the use of different methods and procedures.

\subsection{Stellar extinction}

\begin{figure}
\centerline{\includegraphics[scale=1.70]{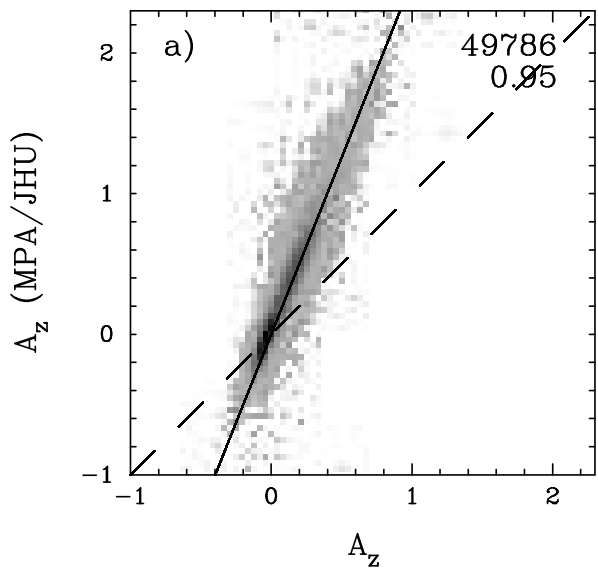}}
\caption{Comparison of the stellar extinction in the $z$-band
determined by the MPA/JHU group with that estimated though the method
described in this paper. The solid line shows a $y = a x$ linear fit
to the data, while the dash ones is the identity line ($y = x$).}
\label{fig:comparisons_AV}
\end{figure}

The MPA/JHU group estimates the $z$-band stellar extinction $A_z$
through the difference between model and measured colours, assuming an
attenuation curve proportional to $\lambda^{-0.7}$. In our case, the
extinction $A_V$ is derived directly from the spectral fitting,
carried out with the Milky Way extinction law (Cardelli \etal 1989,
c.f.\ Section \ref{sec:SynthesisMethod}), for which $A_z = 0.4849 A_V$
assuming $\lambda_z = 8931$ \AA. Fig.~\ref{fig:comparisons_AV} shows
that these two independent estimates are very strongly and linearly
correlated, with a Spearman rank correlation coefficient $r_S =
0.95$. However, the values of $A_z$ reported by the MPA/JHU group
(column 17 of their Stellar Mass Catalogue) are systematically larger
than our values: $A_z({\rm MPA/JHU})\simeq 2.51 A_z({\rm This~Work})$
in the median.

This discrepancy is only apparent. The Galactic extinction law is
substantially harder than $\lambda^{-0.7}$. One thus expects to need
less extinction when modeling a given galaxy with the former law than
with the latter. This was confirmed by STARLIGHT fits to a sub-set of
SDSS galaxies using a $\lambda^{-0.7}$-law, which yield a value of
$A_V$ typically 1.77 times larger than those obtained with the
Cardelli \etal (1989) law. Since the $A_z/A_V$ conversion factors are
0.7030 and 0.4849 for the $\lambda^{-0.7}$ and Cardelli \etal laws,
one finds that the $A_z$ values obtained for the two laws should
differ by a factor of $1.77 \times 0.7030/0.4849 = 2.57$. This is very
close to the empirically derived factor of 2.51
(Fig.~\ref{fig:comparisons_AV}).  We thus conclude that there are no
substantial differences between the MPA/JHU and our estimates of the
stellar extinction other than those implied by the differences in the
reddening laws adopted in the two studies.

\subsection{Stellar masses}

\begin{figure*}
\centerline{\includegraphics[scale=1.20]{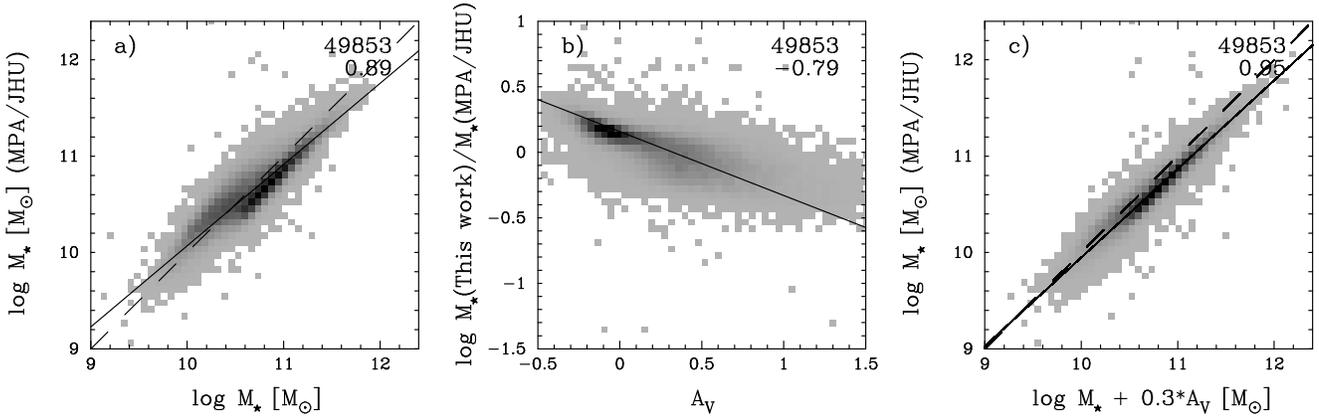}}
\caption{(a) Comparison of the stellar mass determined by the MPA/JHU
group with our estimate.  The solid line shows a $y = a x + b$ linear
fit to the data, while the dash ones is the identity line ($y =
x$). (b) Relation of the ratio between our and the MPA/JHU masses and
our estimate of the V-band stellar extinction. (c) Comparison of the
stellar masses after correcting for differences induced by the
different extinction laws.}
\label{fig:comparisons_Masses}
\end{figure*}

Fig.~\ref{fig:comparisons_Masses}a compares our results for the total
stellar masses to the MPA/JHU extinction-corrected stellar masses
(column 9 of their Stellar Mass Catalogue). The two estimates of
$M_\star$ correlate very well, with $r_S = 0.89$. The quantitative
agreement is also good, with a median difference of just 0.1 dex.
This small offset cannot be attributed to the different IMFs employed
in the two studies (Chabrier 2003 here and Kroupa 2001 in K03), since,
as illustrated in figure 4 of BC03, these two IMFs yield practically
identical $M/L$ ratios. Instead, this offset seems to be due to a
subtle technicality. Whereas we adopt the $M/L$ ratio of the best
$\chi^2$ model, the MPA/JHU group derives $M/L$ comparing the observed
values of the $D_n(4000)$ and $H\delta_A$ indices with a library of
32000 models. Each model is then weighted by its likelihood, and a
probability distribution for $M/L$ is computed. The MPA/JHU mass is
the median of this distribution, which is not necessarily the same as
the best-$\chi^2$ value.  In fact, the Stellar Mass Catalogue of
Brinchmann \etal (2004) lists in its column 8 the best $\chi^2$
masses, which are systematically larger than the median ones by $\sim
0.1$ dex, identical to the median difference identified above.

Part of the scatter in Fig.~\ref{fig:comparisons_Masses}a can be
attributed to the different extinction laws. Extinction contributes $+
0.4 A_{\lambda_0}$ to the estimated $\log M_\star$.  As discussed
above, using a $\lambda^{-0.7}$ law in the synthesis yields values of
$A_V$ which are 1.77 times larger than using the Cardelli \etal (1989)
law.  Furthermore, the mass-to-light ratios obtained with the two laws
are very similar.  From this we expect that $\log M_\star({\rm
This~Work})/ M_\star(\rm MPA/JHU) \approx - 0.3 A_V({\rm This~Work})$,
in good agreement with the observed relation
(Fig.~\ref{fig:comparisons_Masses}b). Correcting for this effect by
adding $0.3 A_V$ to our mass estimates indeed produces a better
correlation, with $r_S = 0.95$, as illustrated in
Fig.~\ref{fig:comparisons_Masses}c. 

Overall, we conclude that the two mass estimates agree to within 0.4
dex. This level of agreement is similar to that recently found by
Drory, Bender \& Hopp (2004) in their comparison between the MPA/JHU
masses and estimates based on SDSS plus 2MASS photometry.  It is quite
remarkable that, despite the substantial differences between our
approaches and the underlying assumptions, the estimated stellar
masses are so similar over a wide range of masses.  On the other hand,
this may not be so surprising given that the MPA/JHU group estimates
of physical properties are ultimately based on a comparison of
observed indices with an extensive library of galaxy spectra
constructed out of the BC03 evolutionary synthesis models (K03). Since
this library spans a wide of metallicities and star-formation
histories, the agreement between our estimates and those of the
MPA/JHU group may be simply indicating that the $N_\star = 45$ SSPs
from BC03 used in our spectral synthesis span a similar
parameter-space to that covered by the K03 models.

\subsection{Velocity dispersion}

\begin{figure}
\centerline{\includegraphics[scale=0.90]{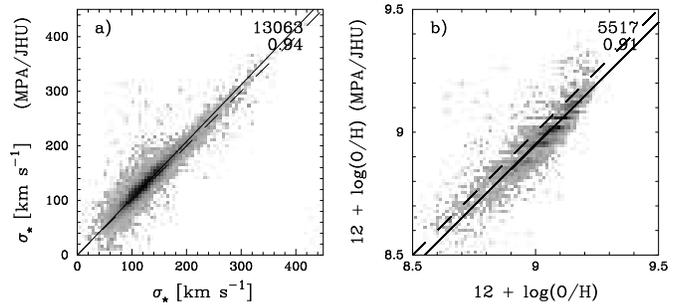}}
\caption{Comparison of the velocity dispersions (a) and oxygen
abundances (b) determined by the MPA/JHU group with our estimates.
The solid line shows a $y = a x + b$ linear fit to the data, while the
dash ones is the identity line ($y = x$).}
\label{fig:comparisons_vd_and_Zneb}
\end{figure}

We use the sub-sample of galaxies with active nuclei to compare our
measurements of absorption line broadening due to galaxy velocity
dispersion and/or rotation, $\sigma_\star$, with those of the MPA/JHU
group, since they list this quantity only in their AGN Catalogue (in
column 16). The comparison is displayed in
Fig.~\ref{fig:comparisons_vd_and_Zneb}a.  The Spearman
correlation-coefficient in this case is $r_S=0.91$ and the median of
the difference between the two estimates is just 9 km$\,$s$^{-1}$,
indicating an excellent agreement between both studies.

\subsection{Emission lines and nebular metallicities}

\label{sec:Comparison_EmissionLines}

Brinchmann \etal (2004) also provide, in their Emission line
Catalogue, emission lines fluxes and equivalent widths which can be
compared with our own measurements.  In both studies the emission
lines are measured after subtracting from the observed spectrum a
model spectrum representing the stellar emission. In our case this is
done with our synthesized spectra.  The MPA/JHU group adopts a similar
approach (see Tremonti \etal 2004 for a brief description) by fitting
the observed continuum with BC03 models. They adopt, however a single
metallicity model and a different extinction law. We have compared the
fluxes and equivalent widths of H$\alpha$, \nii$\lambda6584$,
\oii$\lambda3727$, H$\beta$ and \oiii$\lambda5007$ as measured by our
code and that obtained by the MPA/JHU group. We do not find any
significant difference between these values; the largest discrepancy
($\sim$ 5 per cent) was found for the equivalent widths of H$\alpha$
and \nii, probably due to different estimates of the continuum level
and the associated underlying stellar absorption. Our emission line
measurements are also in good agreement with those in Stasi\'nska
\etal (2004), who fit the Balmer lines with emission and absorption
components, instead of subtracting a starlight model.

In Fig.~\ref{fig:comparisons_vd_and_Zneb}b we plot our estimates of
the nebular oxygen abundance against those obtained by the MPA/JHU
group (Catalogue of Gas Phase Metallicities, median values; see
Tremonti \etal 2004). In order not to introduce any bias due to the
use of different indicators of the oxygen abundance, we have estimated
O/H using the calibration of the (\oii$\lambda$3726,3729 +
\oiii$\lambda$4959,5007)/ H$\beta$ ratio given by Tremonti \etal
(2004) in their equation (1). This calibration is based on
simultaneous fits of the most prominent emission lines with a model
designed for the interpretation of integrated galaxy spectra (Charlot
\& Longhetti 2001). The oxygen abundances estimated in such a way
differ by just $\sim 0.05$ dex, as shown in
Fig.~\ref{fig:comparisons_vd_and_Zneb}b.

Overall, we conclude that our spectral synthesis method yields
estimates of physical parameters in good agreement with those obtained
by the MPA/JHU group, considering the important differences in
approach and underlying assumptions.

\section{Empirical relations}

\label{sec:Correlations}

Yet another way to assess the validity of physical properties derived
through a spectral synthesis analysis is to investigate whether this
method yields astrophysically reasonable results. In this section we
follow this empirical line of reasoning by comparing some results
obtained from our synthesis of SDSS galaxies (which excludes emission
lines) with those obtained from a direct analysis of the emission
lines.  Our aim is to demonstrate that our synthesis results do indeed
make sense.  A more detailed discussion of most of the points below
will be presented in other papers of this series.

\subsection{Stellar and Nebular Metallicities}

\label{sec:Stellar_X_Nebular_Metallicities}

Our spectral synthesis approach yields estimates of the mean
metallicity of the stars in a galaxy, $<Z_\star>$. The analysis of
emission lines, on the other hand, gives estimates of the present-day
abundances in the warm interstellar medium. Although stellar and
nebular metallicities are not expected to be equal, it is reasonable
to expect that they should roughly scale with each other.

\begin{figure}
\centerline{\includegraphics[scale=1.7]{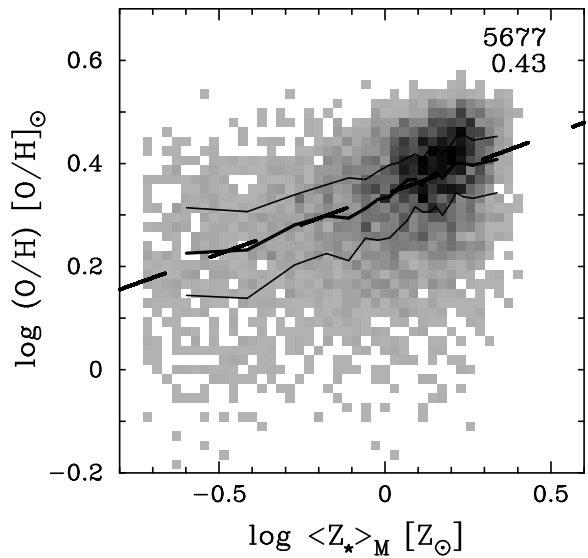}}
\caption{Nebular oxygen abundance versus mass-weighted stellar
metallicity, both in solar units, for normal star-forming galaxies in
our sample.  The median values and quartiles in bins of same number of
objects are shown as thin solid lines.  The dashed line is a robust
fit for the relation.}
\label{fig:zneb_zst}
\end{figure}

Fig.~\ref{fig:zneb_zst} shows the correlation between mass-weighted
stellar metallicities and the nebular oxygen abundance (computed as in
Section \ref{sec:Comparison_EmissionLines}), both in solar
units\footnote{The solar unit adopted for the nebular oxygen abundance
is $12+\log (\rmn{O/H})_\odot=8.69$ (Allende Prieto, Lambert \&
Apslund 2001).}, for our sample of normal star-forming galaxies. A
correlation is clearly seen, although with large scatter ($r_S =
0.42$).  Galaxies with large stellar metallicities also have large
nebular oxygen abundances; galaxies with low stellar abundances tend
to have smaller abundances. The observed scatter is qualitatively
expected due to variations in enrichment histories among galaxies.  A
robust linear fitting gives the following relation:

\begin{equation}
\log \left( \frac{\rmn{O/H}}{0.00049}\right) = 0.34 +
  0.23 \log \left( \frac{<Z_\star>_M}{0.02}\right) 
\end{equation}

\ni with a dispersion of 0.08 dex. Notice that in this expression both
stellar and nebular metallicities are normalized to solar units
($Z_\odot = 0.02$ and $(\rmn{O/H})_\odot=0.00049$ respectively).

Nebular and stellar metallicities are estimated through completely
different and independent methods, so the correlation depicted in
Fig.~\ref{fig:zneb_zst} provides an {\it a posteriori} empirical
validation for the stellar metallicity derived by the spectral
synthesis.  The possibility to estimate stellar metallicities for so
many galaxies is one of the major virtues of spectral synthesis, as it
opens an important window to study the chemical evolution of galaxies
and of the universe as a whole (Panter \etal 2004; Sodr\'e \etal in
prep.).

\subsection{Stellar and Nebular extinctions}

The stellar extinction in the V-band is one of the products of our
STARLIGHT code. A more traditional and completely independent method
to evaluate the extinction consists of comparing the observed
H$\alpha$/H$\beta$ Balmer decrement to the theoretical value.  The
intrinsic value of ${F({\rm H}\alpha)/F({\rm H}\beta)}$ is not very
sensitive to the physical conditions of the gas, ranging from 3.03 for
a gas temperature of 5000 K to 2.74 at 20000 K (Osterbrock
1989). Adopting a value of 2.86 for this ratio, the ``Balmer
extinction'' (Stasi\'nska \etal 2004) is given by

\begin{equation}
A_V^{\rm Balmer} = 6.31 \log \left[
\frac{F({\rm H}\alpha)/F({\rm
H}\beta)}{2.86} \right]
\end{equation}

\ni where the 6.31 coefficient comes from assuming the Cardelli \etal
(1989) extinction curve.

\begin{figure}
\centerline{\includegraphics[scale=1.7]{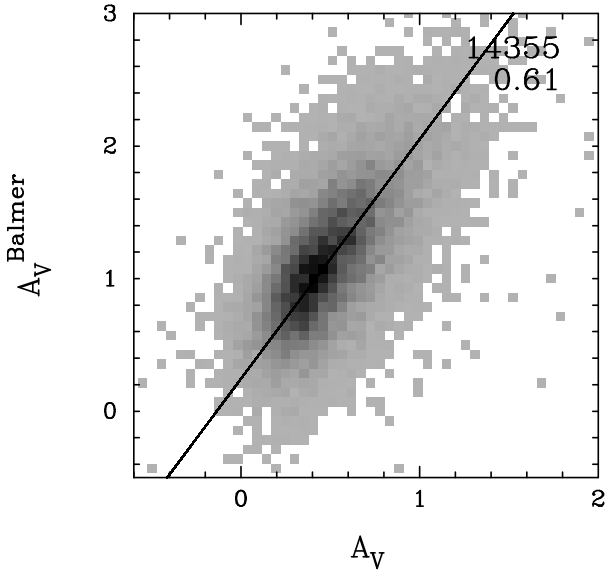}}
\caption{Relation between the nebular ($A_V^{\rm{Balmer}}$) and
stellar ($A_V$) extinctions for our sample of normal star-forming galaxies.
The solid line is a robust fit for the relation.}
\label{fig:AV_Chb}
\end{figure}

Fig.~\ref{fig:AV_Chb} presents a comparison between the stellar $A_V$
and $A_V^{\rm Balmer}$.  These two extinctions are determined in
completely independent ways, and yet, our results show that they are
closely linked, with $r_S = 0.61$. A linear bisector fitting yields

\begin{equation}
\label{eq:AVBalmer_X_AVStellar}
A_V^{\rm Balmer}  = 0.24 + 1.81 A_V
\end{equation}

\ni Note that the angular coefficient in this relation indicates that
nebular photons are roughly twice as extincted as the starlight.  This
``differential extinction'' is in very good qualitative and
quantitative agreement with empirical studies (Fanelli \etal 1988;
Calzetti, Kinney \& Storchi-Bergmann 1994; Gordon \etal 1997;
Mas-Hesse \& Kunth 1999). We shall explore the implications of this
result for the intrinsic colours of star-forming galaxies in another
paper of this series (Stasi\'nska \etal, in prep.).

\subsection{Relations with mean stellar age}

\begin{figure}
\centerline{\includegraphics[scale=0.9]{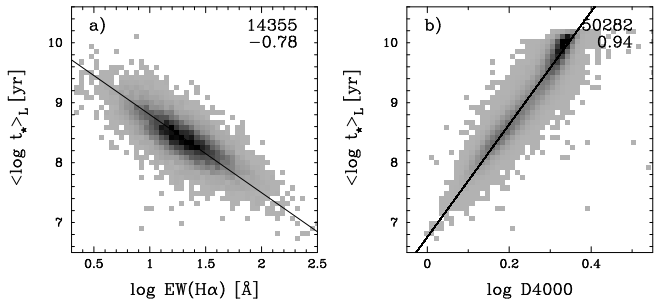}}
\caption{(a) Equivalent width of H$\alpha$ versus the light-weighted
mean stellar age for normal emission line galaxies in our sample.  (b)
Relation between the D4000 index and the light-weighted mean
stellar age. The solid lines are robust fits for the relations.}
\label{fig:age_EWHalpha_D4000}
\end{figure}

The equivalent width (EW) of H$\alpha$ is related to the ratio of
present to past star formation rate of a galaxy (e.g., Kennicutt
1998).  It is thus expected to be smaller for older galaxies.
Fig.~\ref{fig:age_EWHalpha_D4000}a shows the relation between
EW(H$\alpha$) and the mean light-weighted stellar age obtained by our
spectral synthesis. The anti-correlation, which has $r_S = -0.78$, is
evident. From this plot we can derive an empirical relation which can
be used to estimate $<\log t_\star>_L$ (for $\lambda_0 = 4020$ \AA) of
star-forming galaxies through the measurement of EW(H$\alpha$):

\begin{equation}
<\log t_\star>_L  = 10.10 - 1.30 \log \rmn{EW}(H\alpha)
\end{equation}

\ni for $t_\star$ in yr and EW(H$\alpha$) in \AA. It is worth
stressing that these two quantities are obtained independently, since
the spectral synthesis does not include emission lines.

Another quantity that is considered a good age indicator, even for
galaxies without emission lines, is the 4000 \AA\ break, D4000.  We
measured this index following Bruzual (1983), who define D4000 as the
ratio between the average value of $F_\nu$ in the 4050--4250 and
3750--3950 \AA\ bands.  The relation between $<\log t_\star>_L$ and
D4000 is shown in Fig.~\ref{fig:age_EWHalpha_D4000}b. Note that the
concentration of points at the high age end reflects the upper age
limit of the base adopted here, 13 Gyr (c.f.\ Section
\ref{sec:Simulations}).  The correlation is very strong ($r_S =
0.94$), showing that indeed D4000 can be used to estimate empirically
mean light-weighted galaxy ages, despite its metallicity dependence
for very old stellar populations (older than 1 Gyr, as shown by
K03). From the tight relation between $<\log t_\star>_L$ and D4000 we
derive the following empirical relation

\begin{equation}
<\log t_\star>_L = 6.76 + 9.41 \log {\rmn D}4000
\end{equation}

\ni where $t_\star$ is in yr. This robust fit reproduces $<\log
t_\star>_L$ to within an rms dispersion of 0.15 dex.

\subsection{$M_\star$--$\sigma_\star$ relation}

\label{sec:mass-sigma}

\begin{figure}
\centerline{\includegraphics[scale=1.7]{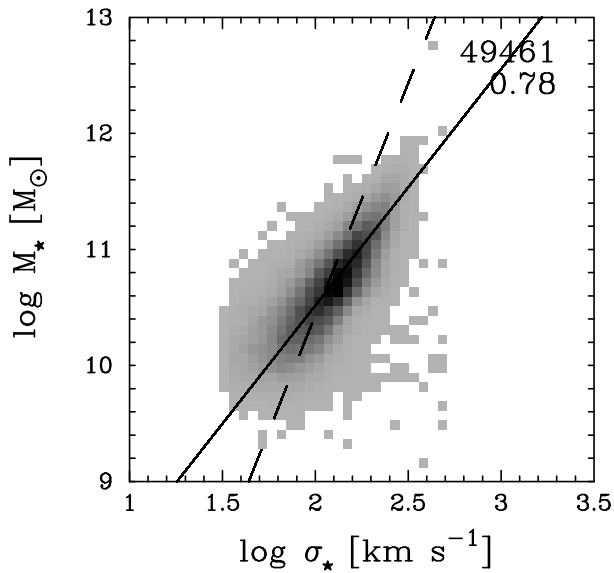}}
\caption{Mass-velocity dispersion relation for our sample. The solid
line is a robust fit for the data while dashed one is a fitting
assuming $M_\star \propto \sigma^4$.}
\label{fig:mass-sigma}
\end{figure}

Fig. \ref{fig:mass-sigma} shows the relation between stellar mass and
$\sigma_\star$, obtained from the synthesis.  The relation is quite
good, with $r_S=0.79$. The solid line displayed in the figure is

\begin{equation}
\log M_\star = 6.44 + 2.04 \log \sigma_\star
\end{equation}

\ni for $M_\star$ in $M_\odot$ and $\sigma_\star$ in km$\,$s$^{-1}$,
obtained with a bisector fitting. The figure also shows as a dashed
line a fit assuming $M_\star \propto \sigma_\star^4$, expected from
the virial theorem under the (unrealistic) assumption of constant mass
surface density. In both cases we have excluded from the fit galaxies
with $\sigma_\star < 35$ km$\,$s$^{-1}$, which corresponds to less
than half the spectral resolution of both data and models.

This is another relation that is expected {\it a priori} if we have in
mind the Faber-Jackson relation for ellipticals and the Tully-Fisher
relation for spirals. For early-type galaxies, $\sigma_\star$ is a
measure of the central velocity dispersion, which is directly linked
to the gravitational potential depth, and, through the virial theorem,
to galactic mass. For late-type systems, $\sigma_\star$ has
contributions of isotropic motions in the bulges, as well as of the
rotation of the disks, and is also expected to relate with galactic
mass. Another aspect that it is interesting to point out in
Fig.~\ref{fig:mass-sigma} is that the dispersion in the
$M_\star$--$\sigma_\star$ relation decreases as we go from
low-luminosity, rotation-dominated systems, for which the values of
$\sigma_\star$ depend on galaxy inclination and bulge-to-disk ratio,
to high-luminosity, mostly early-type systems, which obey a much more
regular (and steeper) relation between $\sigma_\star$ and $M_\star$.

This relation, between a quantity that is not directly linked to the
synthesis, $\sigma_\star$, and another one that is a product of our
synthesis, $M_\star$, is yet another indication that the results of
our STARLIGHT code do make sense.

\section{Summary}

\label{sec:Conclusions}

We have developed and tested a method to fit galaxy spectra with a
combination of spectra of individual simple stellar populations
generated with state-of-the art evolutionary synthesis models. The
main goal of this investigation was to examine the reliability of
physical properties derived in this way. This goal was pursued by
three different means: simulations, comparison with independent
studies, and analysis of empirical results. Our main results can be
summarized as follows:

\begin{enumerate}

\item Simulations tailored to match the characteristics of SDSS
spectra show that the individual SSP strengths, encoded in the
population vector $\vec{x}$, are subjected to large uncertainties, but
robust results can be obtained by compressing $\vec{x}$ into coarser
but useful indices. In particular, physically motivated indices such
as mean stellar ages and metallicities are found to be well recovered
by spectral synthesis even for relatively noisy spectra. Stellar
masses, velocity dispersion and extinction are also found to be
accurately retrieved.

\item We have applied our STARLIGHT code to a volume limited sample of
over 50000 galaxies from the SDSS Data Release 2. The spectral fits
are generally very good, and allow accurate measurements of emission
lines from the starlight subtracted spectrum.  Catalogues of physical
and emission line properties derived for this sample will be made
available in due course. We also report that work is underway to
produce a flexible, user friendly and publicly available version of
STARLIGHT.

\item We have compared our results to those obtained by the MPA/JHU
group (K03; Brinchmann \etal 2004) with a different method to
characterize the stellar populations of SDSS galaxies. The stellar
extinctions and masses derived in these two studies are very strongly
correlated. Furthermore, differences in the values of $A_V$ and
$M_\star$ are found to be mostly due to the differences in the model
ingredients (extinction law).  Our estimates of stellar velocity
dispersions and emission line properties are also in good agreement
with those of the MPA/JHU group.

\item The confidence in the method is further strengthened by several
empirical correlations between synthesis results and independent
quantities. We find strong correlations between stellar and nebular
metallicites, stellar and nebular extinctions, mean stellar age and
the equivalent width of H$\alpha$, mean stellar age and the
4000 \AA\ break, stellar mass and velocity dispersion. These are all
astrophysically reasonable results, which reinforce the conclusion
that spectral synthesis is capable of producing reliable estimates of
physical properties of galaxies.

\end{enumerate}

Overall, these results validate spectral synthesis as a powerful tool
to study the history of galaxies. Other papers in this series will
take advantage of this tool to address issues regarding aspects of
galaxy formation and evolution.

\section*{Acknowledgments}

We thank the organizers of the Guillermo Haro workshop 2004 at the
Instituto Nacional de Astronomia, \'Optica y Electr\'onica (Puebla,
Mexico) for having allowed us to work in a very pleasant and
stimulating environment and we thank the participants for many useful
discussions. We are also in debt with the anonymous referee for
her/his comments and helpful suggestions.  Partial support from CNPq,
FAPESP and the France-Brazil PICS program are also acknowledged.  Last
but not least, we wish to thank G. Bruzual, S. Charlot, and the SDSS
team for their dedication to projects which made the present work
possible.

The Sloan Digital Sky Survey is a joint project of The University of
Chicago, Fermilab, the Institute for Advanced Study, the Japan
Participation Group, the Johns Hopkins University, the Los Alamos
National Laboratory, the Max-Planck-Institute for Astronomy (MPIA),
the Max-Planck-Institute for Astrophysics (MPA), New Mexico State
University, Princeton University, the United States Naval Observatory,
and the University of Washington.  Funding for the project has been
provided by the Alfred P. Sloan Foundation, the Participating
Institutions, the National Aeronautics and Space Administration, the
National Science Foundation, the U.S. Department of Energy, the
Japanese Monbukagakusho, and the Max Planck Society.

\end{document}